\documentclass[aps,prb,nobibnotes,twocolumn,superscriptaddress,bibliography]{revtex4-2}

\usepackage{graphicx}
\usepackage{dcolumn}
\usepackage{xcolor}
\usepackage{amsfonts}
\usepackage{mathrsfs}
\usepackage{amsmath}
\usepackage{color}
\usepackage{amssymb}
\usepackage{xspace}
\usepackage{epstopdf}
\usepackage{longtable}
\usepackage{multirow}
\usepackage{float}
\usepackage{comment}
\usepackage{tabularx}
\usepackage{dsfont}
\usepackage{fontenc}
\usepackage{verbatim}
\usepackage{natbib}
\usepackage{bm}
\usepackage{bbding}
\usepackage{booktabs}
\usepackage{placeins}
\usepackage[colorlinks=true,  pdfstartview=FitV, linkcolor=blue, citecolor=blue, urlcolor=blue]{hyperref}

\begin{document}
\title{Orbital origin of fourfold anisotropic magnetoresistance in Dirac
materials}
\author{Daifeng Tu}
\affiliation{Anhui Provincial Key Laboratory of Low-Energy Quantum Materials and
Devices, High Magnetic Field Laboratory, HFIPS, Chinese Academy of
Sciences, Hefei, Anhui 230031, China}
\affiliation{University of Science and Technology of China, Hefei, 230026, P. R. China}
\author{Can Wang}
\affiliation{Anhui Provincial Key Laboratory of Low-Energy Quantum Materials and
Devices, High Magnetic Field Laboratory, HFIPS, Chinese Academy of
Sciences, Hefei, Anhui 230031, China}
\affiliation{University of Science and Technology of China, Hefei, 230026, P. R. China}
\author{Jianhui Zhou}
\email{jhzhou@hmfl.ac.cn}

\affiliation{Anhui Provincial Key Laboratory of Low-Energy Quantum Materials and
Devices, High Magnetic Field Laboratory, HFIPS, Chinese Academy of
Sciences, Hefei, Anhui 230031, China}
\date{\today}
\begin{abstract}
Fourfold anisotropic magnetoresistance (AMR) have been widely observed
in quantum materials, but the underlying mechanisms remain poorly
understood. Here we find, in a variety of three-dimensional Dirac
materials that can be unifiedly described by the massive Dirac equation,
the intrinsic orbital magnetic moment of electrons vary synchronously
with the magnetic field and give rise to a $\pi$ periodic correction
to its velocity, further leading to unusual fourfold AMR, dubbed orbital
fourfold AMR. Our theory not only explains the observation of fourfold
AMR in bismuth but also uncovers the nature of the dominant fourfold
AMR in thin films of antiferromagnetic topological insulator $\mathrm{MnBi_{2}Te_{4}}$,
which arises from the near cancellation of the twofold AMR from the
surface states and bulk states due to distinct spin-momentum lockings.
Our work provides a new mechanism for creation and manipulation of
orbital fourfold AMR in both conventional conductors and various topological
insulators. 
\end{abstract}
\maketitle
\textit{Introduction.-{}-}Anisotropic magnetoresistance (AMR), a fundamental
phenomenon in magnetic materials, usually arises from the interaction
between the electrons and magnetizations and has many useful functionality
in magnetic sensors and data recording technologies \citep{Ejsing2004APL,Bason2006JAP,Chappert2007NM}.
Previous studies for noncrystalline and crystalline materials showed
that both AMR and the related planar Hall effect (PHE) are $\pi$-periodic
(twofold symmetric component) in the angle between the direction of
electric current and magnetic field \citep{Tang2003PRL,Ritzinger2023RSOS}.
Notably, fourfold AMR was observed in a wide variety of materials
\citep{vanGorkom2001PRB,Muduli2005PRB,Limmer2006PRB,Rushforth2007PRL,Bason2009PRB,Naftalis2011PRB,Annadi2013PRB,DingZ2013jap,xiao2015jap2,Oogane2018jjap,Ahadi2019PRB,ZengFL2020PRL,Wadehra2020NC,Battilomo2021PRB,DaiY2022PRL,Song2022NC},
such as ferromagnetic CoFe alloys \citep{ZengFL2020PRL}, FePt epitaxial
films \citep{DaiY2022PRL} and antiferromagnetic $\mathrm{EuTi_{2}O_{3}}$
\citep{Ahadi2019PRB} and $\mathrm{Nd_{2}Ir_{2}O_{7}}$ \citep{Song2022NC},
which has been attributed to anisotropic relaxation time, higher order
perturbation of spin-orbit coupling (SOC) and cluster magnetic multipoles
of spins \citep{Yahagi2020jpsj,DaiY2022PRL,Song2022NC}. 
Note that the twofold part usually corresponds to the non-crystalline term, while the fourfold one originates from the crystal structure in high-quality single crystals or epitaxial materials~\citep{Ritzinger2023RSOS}.
In reality, the fourfold component usually appears as a subordinate correction
and superposes on the profile of twofold part. How to generate and
control the predominant fourfold AMR in quantum materials is far unexplored. 

Three-dimensional (3D) Dirac materials including the conventional
conductors \citep{Zawadzki2017jpcm,Andreev2018PRL}, the topological
insulators (TIs), topological semimetals have attracted much attention
due to the fascinating quantum phenomena and the promising applications
in low-energy cost electronics, spintronics and plasmonics \citep{Hasan2010RMP,QiZhangRMP}.
Dirac materials host Dirac fermions that possess the unique energy
dispersion and nontrivial Berry curvature, facilitating the realization
of novel topological phase of matter, such as the quantum anomalous
Hall effect, axion insulator and the Majorana excitations for promising
topological quantum computations \citep{WengQAHE2015,Chang2023RMP}.
It has been shown that the self-rotation of electronic wave-packet
leads to the orbital magnetic moment (OMM) which shares the same symmetry
properties as the Berry curvature \citep{Xiao2010RMP,Liu2021NRP,atencia2024omm}.
Here, we demonstrate numerous and easily accessible Dirac materials
provide a platform to explore novel electromagnetic response associated
with intrinsic OMM of electrons. Recently, both bismuth \citep{YangSY2020PRR}
and the antiferromagnetic TI $\mathrm{MnBi_{2}Te_{4}}$ \citep{Wu2022NL}
exhibit notable fourfold AMR. For bismuth, the unusual AMR was attributed
to the anisotropic classical orbital magnetoresistance together with
the chiral anomaly scenario \citep{YangSY2020PRR}, while the anomalous
angular-dependence in AMR/PHE was ascribed to the field-dependent
carrier densities in strong fields \citep{Yamada2021PRB}. Both bismuth
and $\mathrm{MnBi_{2}Te_{4}}$ could be described by massive Dirac
equation but the unified mechanism is still lacking, in particular
the role of Dirac surface states. 

In this Letter, we show the fourfold AMR originates from the intrinsic
OMM of electrons via modifying the velocity in 3D Dirac materials.
We can quantitatively explain the fourfold AMR and its anomalous evolution
of AMR with the magnetic fields in both bismuth and $\mathrm{MnBi_{2}Te_{4}}$.
It has been shown that the twofold AMR of the surface states in TIs
could cancel with that of the bulk states due to the distinct spin-momentum
locking, leading to a dominant fourfold part. In addition, the fourfold
AMR could appear in $\mathrm{Bi_{2-x}Sb_{x}Te_{3}}$ through tuning
the carrier density by chemical doping and electrical gating. 

\textit{\textcolor{black}{Formalism.-{}-}}To investigate the novel
phenomena associated with OMM in 3D Dirac materials within the semiclassical
regime, we start from the the semiclassical dynamics of the electronic
wave packet, which includes both the Berry curvature and OMM \citep{Sundaram1999PRB,Xiao2010RMP}
\begin{equation}
\dot{\mathbf{r}}=\frac{1}{\hbar}\frac{\partial\bar{\varepsilon}_{n\mathbf{k}}}{\partial\mathbf{k}}-\dot{\mathbf{k}}\times\mathbf{\Omega}_{n\mathbf{k}},\,\hbar\dot{\mathbf{k}}=-e(\mathbf{E}+\dot{\mathbf{r}}\times\mathbf{B}),\label{eq:EOM}
\end{equation}
where $\mathbf{r}$ and $\mathbf{k}$ are the position and momentum
of the center of the wave packet of Bloch electrons. It can be seen
that, the Berry curvature $\mathbf{\Omega}_{n\mathbf{k}}=\nabla_{\mathbf{k}}\times\left\langle u_{n\mathbf{k}}\right|i\nabla_{\mathbf{k}}\left|u_{n\mathbf{k}}\right\rangle $
gives rise to an anomalous velocity, where $|u^{n}\rangle$ is the
periodic part of the Bloch wave function. The OMM of state $\left|u_{n\mathbf{k}}\right\rangle $,
$\mathbf{m}_{n\mathbf{k}}=i\left(e/2\hbar\right)\left\langle \nabla_{\mathbf{k}}u_{n\mathbf{k}}\right|\left(\varepsilon_{n\mathbf{k}}-H\left(\mathbf{k}\right)\right)\times\left|\nabla_{\mathbf{k}}u_{n\mathbf{k}}\right\rangle $,
modifies the electron energy as $\bar{\varepsilon}_{n\mathbf{k}}=\varepsilon_{n\mathbf{k}}-\mathbf{m}_{n\mathbf{k}}\cdot\mathbf{B}$,
where $H\left(\mathbf{k}\right)$ is the Bloch Hamiltonian \citep{Xiao2005PRL,Thonhauser2005PRL}.
The OMM can be understood as an additional magnetic moment caused
by self-rotating of the wave packet around its center.

The electric current density is given by 
\begin{equation}
\mathbf{J}=-e\intop[d\mathbf{k}]D^{-1}f_{\mathbf{k}}\mathbf{\dot{\mathbf{r}}},\label{eq:Current}
\end{equation}
where $D=\left[1+\frac{e}{\hbar}(\mathbf{B}\cdot\mathbf{\Omega})\right]^{-1}$
is the modification to the density of states in the phase space. In
the presence of the spatially homogenous external fields, the distribution
function $f_{\mathbf{k}}$ can be determined by the Boltzmann equation
$\dot{\mathbf{k}}\cdot\partial_{\mathbf{k}}f_{\mathbf{k}}=-\frac{f_{\mathbf{k}}-f_{0}}{\tau(\mathbf{k})}$
within the relaxation time approximation \citep{EPhonon2001Ziman},
where $\tau(\mathbf{k})$ is the relaxation time and $f_{0,\mathbf{k}}=\left\{ \mathrm{exp}\left[\frac{1}{kT}(\bar{\varepsilon}_{\mathbf{k}}-\mu)\right]+1\right\} ^{-1}$
is the unperturbed Fermi distribution function. In the linear-response
regime, We have $f_{\mathbf{k}}=f_{0,\mathbf{k}}+f_{1,\mathbf{k}}$
with 
\[
f_{1,\mathbf{k}}=-\tau(\mathbf{k})D\left[e\mathbf{E}\cdot\bar{\mathbf{v}}_{\mathbf{k}}+\frac{e^{2}}{\hbar}(\mathbf{E}\cdot\mathbf{B})(\mathbf{\Omega}\cdot\bar{\mathbf{v}}_{\mathbf{k}})\right]\left(-\frac{\partial f_{0}}{\partial\varepsilon}\right).
\]
with $\bar{\mathbf{v}}_{\mathbf{k}}=\mathbf{v}_{k}-\frac{1}{\hbar}\partial_{\mathbf{k}}(\mathbf{m}_{\mathbf{k}}\cdot\mathbf{B})$
being the velocity of electrons. Substituting $f_{\mathbf{k}}$ and
$\dot{\mathbf{r}}$ into Eq. \eqref{eq:Current} yields the electric
conductivity tensor
\begin{align}
\sigma_{\alpha\beta} & =e^{2}\intop[d\mathbf{k}]\tau(\mathbf{k})D\left(-\frac{\partial f_{0}}{\partial\varepsilon}\right)\left(\bar{v}_{\alpha}\bar{v}_{\beta}+\frac{e}{\hbar}\bar{v}_{\alpha}B_{\beta}(\mathbf{\Omega}\cdot\bar{\mathbf{v}}_{\mathbf{k}})\right.\nonumber \\
 & +\frac{e}{\hbar}\bar{v}_{\beta}B_{\alpha}(\mathbf{\Omega}\cdot\bar{\mathbf{v}}_{\mathbf{k}})\left.+\frac{e^{2}}{\hbar^{2}}B_{\alpha}B_{\beta}(\mathbf{\Omega}\cdot\bar{\mathbf{v}}_{\mathbf{k}})^{2}\right),\label{eq:zx}
\end{align}
which allows us to investigate the impacts of OMM on the transport
properties of electrons, in particular, beyond the quantitative corrections
\citep{DaiX2017PRL,Nandy2018SR}. Note that we have omitted the terms
solely associated with Berry curvature such as the anomalous Hall
effect and the chiral magnetic effect of Weyl fermions as well as
these higher order terms relevant to the positional shift \citep{GaoY2017PRB,Gao2019FoP,HuangYX2023PRL}.
In this work, we focus on the magnetotransport properties on single-particle
level and neglect the role of electron-electron interaction.

\textit{\textcolor{black}{Massive Dirac fermions.-{}-}}Dirac equation
plays a crucial role in understanding and realizing the fascinating
topological phases of matter as well as the novel electromagnetic
responses in a large family of quantum materials, such as TIs, topological
semimetals \citep{shen2013TIDirac,Armitage2018RMP,Lv2021RMP}. We
here utilize the effective Dirac model with a modified mass, which
has been used to describe the topological Anderson localization, negative
MR and resistivity anomaly in 3D topological materials \citep{GuoHM2010PRL,DaiX2017PRL,FuBo2020PRL}
\begin{equation}
H_{0}(\mathbf{k})=\left(\begin{array}{cccc}
\mathcal{M}(\mathbf{k}) & 0 & Ak_{z} & Ak_{-}\\
0 & \mathcal{M}(\mathbf{k}) & Ak_{+} & -Ak_{z}\\
Ak_{z} & Ak_{-} & \mathcal{-M}(\mathbf{k}) & 0\\
Ak_{+} & -Ak_{z} & 0 & -\mathcal{M}(\mathbf{k})
\end{array}\right),
\end{equation}
where $\mathcal{M}(\mathbf{k})=M-Fk^{2}$ is the gap, and $k_{\pm}=k_{x}\pm ik_{y}$.
Both the conduction and valence bands are doubly degenerate $\varepsilon_{\mathbf{k},0,\pm}=\pm\varepsilon$
with $\varepsilon=\sqrt{A^{2}k^{2}+\mathcal{M}^{2}\left(\mathbf{k}\right)}$.
The two wave functions of electrons in the conduction bands are $|u_{1,2}(\mathbf{k})\rangle$.
Some cumbersome calculations give us the $\mathrm{SU}(2)$ Berry connection
$\boldsymbol{\mathscr{\mathcal{A}}}=A^{2}\mathbf{k}\times\boldsymbol{\sigma}/2\varepsilon\left[\varepsilon-\mathcal{M}\left(\mathbf{k}\right)\right]$
and the corresponding Berry curvature 
\begin{align}
\varOmega_{i} & =\frac{A^{2}}{2\varepsilon^{3}}\left\{ (\mathcal{M}+2Fk^{2})\sigma_{i}+\left[A^{2}(\mathcal{M}-\varepsilon-2Fk^{2})\right.\right.\nonumber \\
 & \left.\left.+4F\mathcal{M}(\varepsilon-\mathcal{M})\right]\frac{(\mathbf{k}\cdot\boldsymbol{\sigma})k_{i}}{(\varepsilon-\mathcal{M})^{2}}\right\} ,
\end{align}
Noted that the OMM and Berry curvature here satisfy the relation of
$\mathbf{m}=\frac{e}{\hbar}\varepsilon\mathbf{\varOmega}$ due to
the particle-hole symmetry, as the standard Dirac equation \citep{Chang2008jpcm}. 

To reveal the impacts of the magnetic field on transport properties
in the planar Hall geometry, we consider a Zeeman term 
\begin{equation}
H_{z}=\frac{\mu_{B}}{2}\left(\begin{array}{cccc}
g_{\perp}B_{z} & g_{\parallel}B_{-} & 0 & 0\\
g_{\parallel}B_{+} & -g_{\perp}B_{z} & 0 & 0\\
0 & 0 & g_{\perp}B_{z} & g_{\parallel}B_{-}\\
0 & 0 & g_{\parallel}B_{+} & -g_{\perp}B_{z}
\end{array}\right),
\end{equation}
where $B_{\pm}=B_{x}\pm iB_{y}$, $g_{\perp/\parallel}$ are Lande
g factor and $\mu_{B}$ is the Bohr magneton. $H_{z}$ breaks the
SU(2) symmetry of electrons in the conduction and valance bands as
well as the relevant Berry curvature and OMM. 

In order to gain clear insights into the impacts of OMM on AMR, we
consider an isotropic g-factor $g_{\perp}=g_{\parallel}=g$ and a
constant mass $\mathcal{M}\left(\mathbf{k}\right)=M$. The corresponding
eigenstates can be solved analytically, and the energy of the top
conduction band is modified to $\varepsilon_{+,\uparrow}(\mathbf{k})=\sqrt{\varepsilon_{k,0}^{2}+\widetilde{B}^{2}+2|\widetilde{\mathbf{B}}|\sqrt{A^{2}(\mathbf{k}\cdot\mathbf{n})^{2}+M^{2}}}$,
where $\widetilde{\mathbf{B}}=\frac{\mu_{B}}{2}g\mathbf{B}$ is the
effect Zeeman field, $\mathbf{n}=\mathbf{B}/|\mathbf{B}|$ is unit
vector along the direction of the magnetic field. For a magnetic field
along the $z$ direction, the Berry curvature and OMM of conduction
band can be obtained as {[}details are given in the Supplemental Material
(SM) \citep{SMAMR}{]} 
\begin{align}
\mathbf{\Omega}\left(\mathbf{k}\right) & =-\frac{n_{z}A^{4}}{2|B_{z}|\gamma\varepsilon_{+,\uparrow}^{3}}\left(k_{x}k_{z},k_{y}k_{z},\frac{\gamma^{2}+|\widetilde{\mathbf{B}}|\gamma}{A^{2}}\right),
\end{align}
with $\gamma=\sqrt{A^{2}k_{z}^{2}+M^{2}}$. The relevant OMM is given
as $\mathbf{m}=\frac{e}{\hbar}\varepsilon_{+,\uparrow}\mathbf{\Omega}$.
Together with the behavior of Berry curvature and OMM with the magnetic
fields in different directions, one finds that the
direction of Berry curvature/OMM almost synchronously changes along
with the direction of the magnetic field (Fig. 1 in SM~\citep{SMAMR}), leading to a nonzero anomalous
Hall effect $\sigma_{xy}\neq0$. More importantly, the correction
to the energy dispersion from the OMM $\mathbf{m}_{\mathbf{k}}(\mathbf{B})\cdot\mathbf{B}$
depends on the angle of the magnetic field with a period of $\pi$.
As a result, the $\pi/2$-periodic dependence of electric conductivity
on the angle of the magnetic field emerges through the quadratic term
like $\left(\mathbf{m}_{\mathbf{k}}\left(\mathbf{B}\right)\cdot\mathbf{B}\right)^{2}$
in Eq. \eqref{eq:zx}, namely, the intrinsic orbital fourfold AMR. 

There are several salient features of the intrinsic orbital fourfold
AMR here \citep{InEx}. First, it is even in the magnetic field and
odd in the relaxation time, reflecting the magnetoresistance nature
\citep{SHTRIKMAN1965SSC}, in contrast to other in-plane magnetotransport
phenomena \citep{Cullen2021PRL,CaoJ2023PRL,WangH2024PRL}. Second,
this orbital fourfold AMR depends on the form of unique spin-momentum
locking of 3D Dirac electrons and the associated intrinsic OMM but
does not require the extrinsic anisotropic relaxation time \citep{DaiY2022PRL}
or special crystal symmetry for usual ferromagnets \citep{Yahagi2020jpsj}.
Third, unlike the fourfold AMR in conventional magnetic materials,
the permanent magnetic orders from the spin degree of freedom of electrons
is not a perquisite for the present fourfold AMR. Next we would like
to apply our theory to the observed fourfold AMR in two representative
Dirac materials: semimetal bismuth and metallic $\mathrm{MnBi_{2}Te_{4}}$.

\begin{figure}
\includegraphics[scale=0.55]{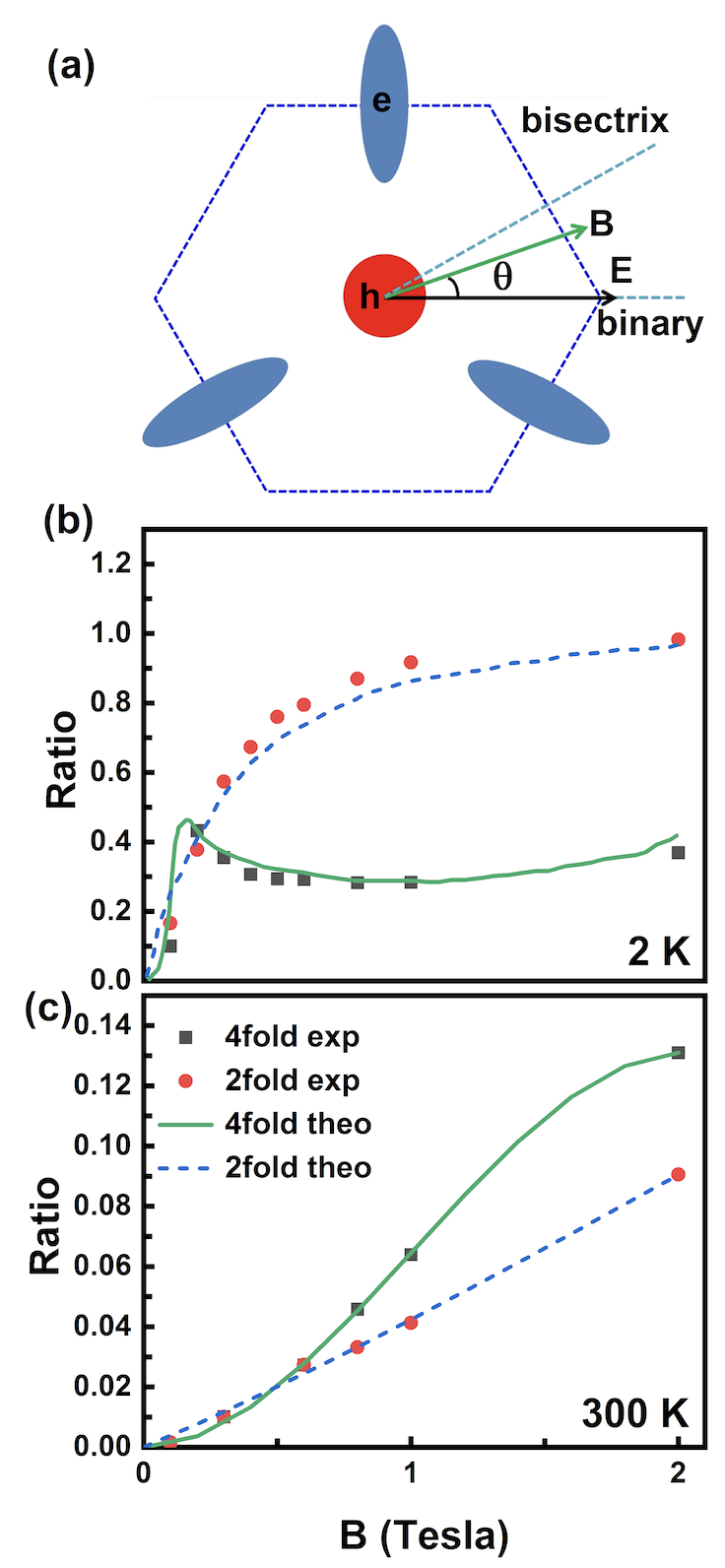}\caption{(a) Schematic illustration of Fermi surface in the binary-bisectrix
plane of Bismuth. Comparison between experimental data $\rho_{11}$
of Bismuth (colored scatters) \citep{dataBi} and theoretical results
of amplitude ratio of AMR (solid and dashed lines) at 2 K (b) and
300 K (c). The orbital fourfold AMR increases as $B^{2}$ for the
weak magnetic field, then decreases as $\left|B\right|^{3}$, and
the position of this peak is affected by temperature. The parameters
in numerical calculations are given \citep{SMAMR}. The dashed lines
are depicted to guide the eye. \label{Bi_AMR}}

\end{figure}

\textit{\textcolor{black}{AMR in bismuth.-{}-}}Bismuth, elemental
semimetal, exhibits many intriguing quantum phenomena \citep{Gygax1986PRL,Collaudin2015PRX,Feldman2016Science,Ito2016PRL,Zhu2018jpcm}
and hosts new higher order topological phases \citep{Schindler2018NP,Aggarwal2021NC}.
Fig. \ref{Bi_AMR}(a) shows that there are three electron pockets
near the equivalent $L$-point and hole pocket near the $T$-point
\citep{LiuY1995PRB}. The electronic states near the $L$-point can
be effectively described by the Dirac-Wolff Hamiltonian and exhibit
small effective mass (about $10^{-3}m_{e}$ with bare electron mass
$m_{e}$) and large anisotropic g-factor (several hundreds) \citep{WOLFF1964jpcs,Fuseya2015PRL},
which make it very sensitive to the magnetic field. Recently, the
fourfold AMR/PHE has been observed when the electric current and the
rotated magnetic field are in the binary-bisectrix plane in micro-thick
(111) thin films of single-crystal bismuth \citep{YangSY2020PRR}.
However, the microscopic mechanism is still under debate \citep{Bi4amr}. 

Here we numerically calculate the AMR with the OMM of Eq. \eqref{eq:zx}, present the results
in Figs. \ref{Bi_AMR}(b, c) and make some comparison with experimental
ones ($\rho_{11}$ from 0.1 T to 2 T in Figs. S5-6 in Ref. \citep{YangSY2020PRR}.
The detailed analysis of $\rho_{12}$ is quite similar and given in
SM \citep{SMAMR}.). In order to further reveal nature of unusual
AMR, we expand $\left[\partial_{\mathbf{k}}(\mathbf{m}_{\mathbf{k}}\left(\mathbf{B}\right)\cdot\mathbf{B})\right]^{2}$
with respect to the magnetic field (see Eq. (S17) in SM \citep{SMAMR})
and find two orders of the expansion of make primary contribution
to the fourfold AMR. Specifically, as shown in Fig. \ref{Bi_AMR}(c),
at 300 K, the twofold AMR can be captured by $a_{1}B$, and the fourfold
AMR can be described by $a_{2}B^{2}+a_{3}\left|B\right|^{3}$, where
$a_{1}$ and $a_{2}$ are positive, but $a_{3}$ is negative. It should
be noted that the peak in the fourfold AMR at 2 K can be understood
from the approximative expansion above. The positive $a_{2}B^{2}$
term dominates at low field, the negative $a_{3}\left|B\right|^{3}$
term becomes important as increase the magnetic field, inevitably
producing the maximum near $B \approx 0.3$~T. The fourfold AMR at 2 K and 300 K primely lie
in the decreasing regime and the increasing regimes due to the thermal
broadening, respectively. The calculations are in quantitative agreement
with the experimental results of the fourfold AMR and the nontrivial
field and temperature dependences in low field regime.

\textit{AMR of surface states in $\mathrm{MnBi_{2}Te_{4}}$.-{}-}As the band inversion occurs,
the material would enter the topological insulating phase, supporting
two-dimensional (2D) Dirac surface states \citep{Bansil2016RMP}.
The effective model for the surface states in the $x-y$ plane can
be written as \citep{zhang2009NP,Shan2010njp,JiangYD2022PRB} 
\begin{equation}
H_{sur}=\hbar v_{F}(k_{x}\sigma_{y}-k_{y}\sigma_{x}),\label{eq:surface}
\end{equation}
which resembles the 2D Rashba SOC $\left(\boldsymbol{k}\times\boldsymbol{\sigma}\right)\cdot z$
\citep{Manchon2015NM}. Here $v_{F}$ is the effective Fermi velocity
of the surface states, $\sigma_{x,y}$ are the Pauli matrices for
the real spin of electrons. Note the Zeeman term of an in-plane magnetic
field, $H_{z}=\frac{1}{2}g\mu_{B}(B_{x}\sigma_{x}+B_{y}\sigma_{y})$,
does not open a gap but only shifts the position of surface Dirac
point \citep{ZhangSB2021PRL}. It should be pointed out that, for
2D systems, the OMM is usually normal to the plane and would not modify
the energy dispersion and velocity of electrons. 

\begin{figure}
\includegraphics[scale=0.6]{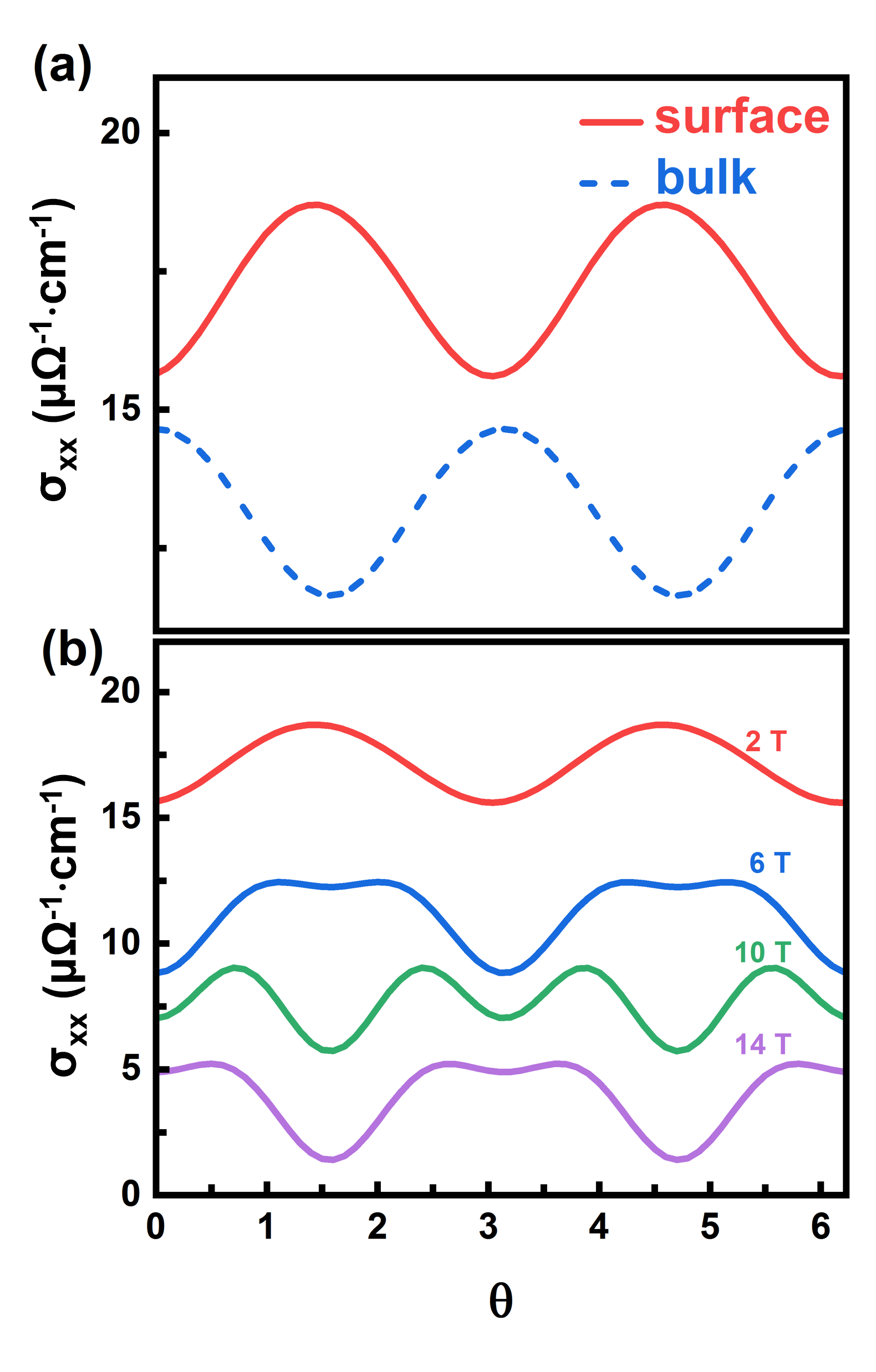}\caption{(a) The orbital twofold AMR of the 2D surface states and the bulk
states in antiferromagnetic TI $\mathrm{MnBi_{2}Te_{4}}$ \citep{Wu2022NL},
which amplitudes have opposite sign. (b) Evolution of AMR with the
magnetic field. As the magnetic field increases, the orbital fourfold
AMR becomes noticeable and then dominates over the twofold part around
10 T. In our calculations, the Fermi energy $E_{F}=500\;\mathrm{meV}$
is measured from the center of the bulk band gap and T= 10 K. Other
parameters are obtained from first-principle calculations \citep{MBTParameters}.
\label{MBT_AMR} }

\end{figure}

To simulate the scatterings between surface Dirac electrons and the
localized magnetic atoms (such as Mn atoms in $\mathrm{MnBi_{2}Te_{4}}$
\citep{Wu2022NL}), we consider the spin-dependent scattering potential
as $U(r)=\mu_{0}\sum_{i}\mathbf{M}\cdot\mathbf{\boldsymbol{\mu}}\delta(\mathbf{r}-\mathbf{R}_{i})$,
similar to the case of topological insulator-ferromagnetic insulator
bilayer \citep{Chiba2017PRB}, where $\mu_{0}$ is the interaction
strength, $\mathbf{\boldsymbol{\mu}}=\frac{1}{2}g\mu_{B}\mathbf{\boldsymbol{\sigma}}$
is spin magnetic moment of electrons, $R_{i}$ is the impurity position,
and $\mathbf{M}$ is the field-polarized magnetic moment of the local
impurities. We can calculate the relaxation time in the Born approximation
and the electric conductivity of the surface states \citep{SMAMR}.
The calculated conductivities for the surface and bulk states are
shown in Fig. \ref{MBT_AMR}(a). The twofold AMR of the surface states
can be understood from the distinction of the spin-momentum lockings.
For the Dirac surface electrons, when the spins of electrons are parallel
or antiparallel to the magnetic field or induced magnetization, the
backscattering is forbidden, the corresponding resistivity will be
much less affected by the magnetic field ($\rho_{\perp}$). On the
other hand, when the spins are perpendicular to the field, the backscattering
becomes allowed due to the broken time reversal symmetry, resulting
in a significant enhancement of resistivity ($\rho_{\parallel}$). That
is a positive anisotropic resistivity (i.e., $\rho_{\parallel}-\rho_{\perp}>0$).
However, for the bulk states with parallel spin and momentum $\boldsymbol{k}\cdot\boldsymbol{\sigma}$,
the direction of field enhanced resistivity and the direction of resistivity
that is less sensitive to magnetic fields get exchanged, leading to
a negative anisotropic resistivity ($\rho_{\parallel}-\rho_{\perp}<0$).
It was also supported by the specific calculations of the anisotropic
resistivity of a 2D slice of bulk states \citep{SMAMR}

\begin{equation}
\sigma_{xx}^{2D}=\frac{8e^{2}\hbar v_{F}^{2}\left[1+(2+\sqrt{3})\cos2\theta\right]}{\sqrt{3}n_{s}(\mu_{0}g\mu_{B})^{2}|\mathbf{M}|^{2}},
\end{equation}
where $n_{s}$ is the density of magnetic impurities and $\theta=\mathrm{arctan}\;(M_{y}/M_{x})$.
One can find that the cancellation of the twofold AMR between the
surface states and the bulk states probably stems from the distinct
spin-momentum lockings, highlighting the role of the interplay between
the magnetic scatterings and SOC in AMR \citep{Vyborny2009PRB}. In
fact, this cancellation mechanism should be generally applicable to
a wide range of  3D TIs including nonmagnetic, magnetic and higher-order
ones \citep{XuYF2019PRL,ZhangDQ2019PRL,ZhangRX2020PRL}. 

When the Fermi level lies in the conduction band of bulk, both the
surface states and the bulk states simultaneously contribute to the
electric transport. Because of the much higher mobility, the surface
states dominates over the bulk ones at lower fields. It is probable
that, under some proper conditions, they can almost cancel out, making
the fourfold AMR dominant. We numerically calculate the AMR that from
both the bulk and the surface states of $\mathrm{MnBi_{2}Te_{4}}$,
as shown in Fig. \ref{MBT_AMR}(b). It can be seen that when the magnetic
field is around 10 T, the orbital fourfold AMR dominates. Further
increasing of the magnetic field, the orbital twofold AMR becomes
dominant again, but the sign of its amplitude changes, which is in
line with the observed AMR \citep{Wu2022NL}. 

Recently, when the Fermi level lies in the bulk gap, the twofold AMR
of Dirac surface states has been observed in the dual-gated devices
of thin films of nonmagnetic TI $\mathrm{Bi}{}_{2}\mathrm{SbTe}_{3}$
\citep{taskin_planar_2017} and been understood in different mechanisms,
such as scatterings of magnetic impurities \citep{taskin_planar_2017}
and tilting of surface Dirac cones with nonlinear momentum terms \citep{ZhengSH2020PRB}.
The electrical gating and chemical doping could tune the carrier density
and shift the Fermi level into the bulk energy band, facilitating
the realization of the predominant orbital fourfold AMR therein. 

\textit{Summary.-{}-}We have found the tunable orbital fourfold AMR
due to OMM of electrons in various 3D Dirac materials. The unique
spin-momentum lockings in topological bulk states and surface states
cause competition between their twofold AMR, leaving predominant fourfold
component. Our calculations are in quantitative agreement with the
fourfold AMR in both bismuth and $\mathrm{MnBi_{2}Te_{4}}$ and the
nontrivial magnetic-field dependence. This work reveals the significance
of intrinsic OMM in novel AMR, inspiring more investigations of intriguing
quantum phenomena associated with OMM in quantum materials, such as
the nonlinear PHE \citep{HeP2019PRL}. Moreover, the first-principles
calculations of OMM in the semiclassical transport theory could provide
us a new and significant ingredient to deep investigate the novel
magnetotransport properties of realistic quantum materials with complicated
Fermi surfaces \citep{ZhangSN2019PRB} and relaxation time \citep{DaiY2022PRL}. 

\textcolor{black}{The authors acknowledge Y. Gao, Q. Niu, T. Qin and
Z. Yuan for insightful discussions. This work was supported by National
Key R\&D Program of the MOST of China (Grant No. 2024YFA1611300) and
the National Natural Science Foundation of China under Grants (No.
U2032164 and No. 12174394)} and by the High Magnetic Field Laboratory
of Anhui Province and by the HFIPS Director\textquoteright s Fund
(Grant Nos. YZJJQY202304 and BJPY2023B05) and Chinese Academy of Sciences
under contract No. JZHKYPT-2021-08.

\bibliography{PHEAMR}

\begin{thebibliography}{84}%
\makeatletter
\providecommand \@ifxundefined [1]{%
 \@ifx{#1\undefined}
}%
\providecommand \@ifnum [1]{%
 \ifnum #1\expandafter \@firstoftwo
 \else \expandafter \@secondoftwo
 \fi
}%
\providecommand \@ifx [1]{%
 \ifx #1\expandafter \@firstoftwo
 \else \expandafter \@secondoftwo
 \fi
}%
\providecommand \natexlab [1]{#1}%
\providecommand \enquote  [1]{``#1''}%
\providecommand \bibnamefont  [1]{#1}%
\providecommand \bibfnamefont [1]{#1}%
\providecommand \citenamefont [1]{#1}%
\providecommand \href@noop [0]{\@secondoftwo}%
\providecommand \href [0]{\begingroup \@sanitize@url \@href}%
\providecommand \@href[1]{\@@startlink{#1}\@@href}%
\providecommand \@@href[1]{\endgroup#1\@@endlink}%
\providecommand \@sanitize@url [0]{\catcode `\\12\catcode `\$12\catcode
  `\&12\catcode `\#12\catcode `\^12\catcode `\_12\catcode `\%12\relax}%
\providecommand \@@startlink[1]{}%
\providecommand \@@endlink[0]{}%
\providecommand \url  [0]{\begingroup\@sanitize@url \@url }%
\providecommand \@url [1]{\endgroup\@href {#1}{\urlprefix }}%
\providecommand \urlprefix  [0]{URL }%
\providecommand \Eprint [0]{\href }%
\providecommand \doibase [0]{https://doi.org/}%
\providecommand \selectlanguage [0]{\@gobble}%
\providecommand \bibinfo  [0]{\@secondoftwo}%
\providecommand \bibfield  [0]{\@secondoftwo}%
\providecommand \translation [1]{[#1]}%
\providecommand \BibitemOpen [0]{}%
\providecommand \bibitemStop [0]{}%
\providecommand \bibitemNoStop [0]{.\EOS\space}%
\providecommand \EOS [0]{\spacefactor3000\relax}%
\providecommand \BibitemShut  [1]{\csname bibitem#1\endcsname}%
\let\auto@bib@innerbib\@empty
%</preamble>
\bibitem [{\citenamefont {Ejsing}\ \emph {et~al.}(2004)\citenamefont {Ejsing},
  \citenamefont {Hansen}, \citenamefont {Menon}, \citenamefont {Ferreira},
  \citenamefont {Graham},\ and\ \citenamefont {Freitas}}]{Ejsing2004APL}%
  \BibitemOpen
  \bibfield  {author} {\bibinfo {author} {\bibfnamefont {L.}~\bibnamefont
  {Ejsing}}, \bibinfo {author} {\bibfnamefont {M.~F.}\ \bibnamefont {Hansen}},
  \bibinfo {author} {\bibfnamefont {A.~K.}\ \bibnamefont {Menon}}, \bibinfo
  {author} {\bibfnamefont {H.~A.}\ \bibnamefont {Ferreira}}, \bibinfo {author}
  {\bibfnamefont {D.~L.}\ \bibnamefont {Graham}},\ and\ \bibinfo {author}
  {\bibfnamefont {P.~P.}\ \bibnamefont {Freitas}},\ }\bibfield  {title}
  {\bibinfo {title} {{Planar Hall effect sensor for magnetic micro- and
  nanobead detection}},\ }\href {https://doi.org/10.1063/1.1759380} {\bibfield
  {journal} {\bibinfo  {journal} {Applied Physics Letters}\ }\textbf {\bibinfo
  {volume} {84}},\ \bibinfo {pages} {4729} (\bibinfo {year}
  {2004})}\BibitemShut {NoStop}%
\bibitem [{\citenamefont {Bason}\ \emph {et~al.}(2006)\citenamefont {Bason},
  \citenamefont {Klein}, \citenamefont {Yau}, \citenamefont {Hong},
  \citenamefont {Hoffman},\ and\ \citenamefont {Ahn}}]{Bason2006JAP}%
  \BibitemOpen
  \bibfield  {author} {\bibinfo {author} {\bibfnamefont {Y.}~\bibnamefont
  {Bason}}, \bibinfo {author} {\bibfnamefont {L.}~\bibnamefont {Klein}},
  \bibinfo {author} {\bibfnamefont {J.-B.}\ \bibnamefont {Yau}}, \bibinfo
  {author} {\bibfnamefont {X.}~\bibnamefont {Hong}}, \bibinfo {author}
  {\bibfnamefont {J.}~\bibnamefont {Hoffman}},\ and\ \bibinfo {author}
  {\bibfnamefont {C.~H.}\ \bibnamefont {Ahn}},\ }\bibfield  {title} {\bibinfo
  {title} {{Planar Hall-effect magnetic random access memory}},\ }\href
  {https://doi.org/10.1063/1.2162824} {\bibfield  {journal} {\bibinfo
  {journal} {Journal of Applied Physics}\ }\textbf {\bibinfo {volume} {99}},\
  \bibinfo {pages} {08R701} (\bibinfo {year} {2006})}\BibitemShut {NoStop}%
\bibitem [{\citenamefont {Chappert}\ \emph {et~al.}(2007)\citenamefont
  {Chappert}, \citenamefont {Fert},\ and\ \citenamefont
  {Van~Dau}}]{Chappert2007NM}%
  \BibitemOpen
  \bibfield  {author} {\bibinfo {author} {\bibfnamefont {C.}~\bibnamefont
  {Chappert}}, \bibinfo {author} {\bibfnamefont {A.}~\bibnamefont {Fert}},\
  and\ \bibinfo {author} {\bibfnamefont {F.~N.}\ \bibnamefont {Van~Dau}},\
  }\bibfield  {title} {\bibinfo {title} {{The emergence of spin electronics in
  data storage}},\ }\href {https://doi.org/10.1038/nmat2024} {\bibfield
  {journal} {\bibinfo  {journal} {Nature Materials}\ }\textbf {\bibinfo
  {volume} {6}},\ \bibinfo {pages} {813} (\bibinfo {year} {2007})}\BibitemShut
  {NoStop}%
\bibitem [{\citenamefont {Tang}\ \emph {et~al.}(2003)\citenamefont {Tang},
  \citenamefont {Kawakami}, \citenamefont {Awschalom},\ and\ \citenamefont
  {Roukes}}]{Tang2003PRL}%
  \BibitemOpen
  \bibfield  {author} {\bibinfo {author} {\bibfnamefont {H.~X.}\ \bibnamefont
  {Tang}}, \bibinfo {author} {\bibfnamefont {R.~K.}\ \bibnamefont {Kawakami}},
  \bibinfo {author} {\bibfnamefont {D.~D.}\ \bibnamefont {Awschalom}},\ and\
  \bibinfo {author} {\bibfnamefont {M.~L.}\ \bibnamefont {Roukes}},\ }\bibfield
   {title} {\bibinfo {title} {{Giant Planar Hall Effect in Epitaxial (Ga,Mn)As
  Devices}},\ }\href {https://doi.org/10.1103/PhysRevLett.90.107201} {\bibfield
   {journal} {\bibinfo  {journal} {Phys. Rev. Lett.}\ }\textbf {\bibinfo
  {volume} {90}},\ \bibinfo {pages} {107201} (\bibinfo {year}
  {2003})}\BibitemShut {NoStop}%
\bibitem [{\citenamefont {Ritzinger}\ and\ \citenamefont
  {V\'yborn\'y}(2023)}]{Ritzinger2023RSOS}%
  \BibitemOpen
  \bibfield  {author} {\bibinfo {author} {\bibfnamefont {P.}~\bibnamefont
  {Ritzinger}}\ and\ \bibinfo {author} {\bibfnamefont {K.}~\bibnamefont
  {V\'yborn\'y}},\ }\bibfield  {title} {\bibinfo {title} {{Anisotropic
  magnetoresistance: materials, models and applications}},\ }\href
  {https://doi.org/10.1098/rsos.230564} {\bibfield  {journal} {\bibinfo
  {journal} {Royal Society Open Science}\ }\textbf {\bibinfo {volume} {10}},\
  \bibinfo {pages} {230564} (\bibinfo {year} {2023})}\BibitemShut {NoStop}%
\bibitem [{\citenamefont {van Gorkom}\ \emph {et~al.}(2001)\citenamefont {van
  Gorkom}, \citenamefont {Caro}, \citenamefont {Klapwijk},\ and\ \citenamefont
  {Radelaar}}]{vanGorkom2001PRB}%
  \BibitemOpen
  \bibfield  {author} {\bibinfo {author} {\bibfnamefont {R.~P.}\ \bibnamefont
  {van Gorkom}}, \bibinfo {author} {\bibfnamefont {J.}~\bibnamefont {Caro}},
  \bibinfo {author} {\bibfnamefont {T.~M.}\ \bibnamefont {Klapwijk}},\ and\
  \bibinfo {author} {\bibfnamefont {S.}~\bibnamefont {Radelaar}},\ }\bibfield
  {title} {\bibinfo {title} {{Temperature and angular dependence of the
  anisotropic magnetoresistance in epitaxial Fe films}},\ }\href
  {https://doi.org/10.1103/PhysRevB.63.134432} {\bibfield  {journal} {\bibinfo
  {journal} {Phys. Rev. B}\ }\textbf {\bibinfo {volume} {63}},\ \bibinfo
  {pages} {134432} (\bibinfo {year} {2001})}\BibitemShut {NoStop}%
\bibitem [{\citenamefont {Muduli}\ \emph {et~al.}(2005)\citenamefont {Muduli},
  \citenamefont {Friedland}, \citenamefont {Herfort}, \citenamefont
  {Sch\"onherr},\ and\ \citenamefont {Ploog}}]{Muduli2005PRB}%
  \BibitemOpen
  \bibfield  {author} {\bibinfo {author} {\bibfnamefont {P.~K.}\ \bibnamefont
  {Muduli}}, \bibinfo {author} {\bibfnamefont {K.-J.}\ \bibnamefont
  {Friedland}}, \bibinfo {author} {\bibfnamefont {J.}~\bibnamefont {Herfort}},
  \bibinfo {author} {\bibfnamefont {H.-P.}\ \bibnamefont {Sch\"onherr}},\ and\
  \bibinfo {author} {\bibfnamefont {K.~H.}\ \bibnamefont {Ploog}},\ }\bibfield
  {title} {\bibinfo {title} {{Antisymmetric contribution to the planar Hall
  effect of ${\mathrm{Fe}}_{3}\mathrm{Si}$ films grown on $\mathrm{GaAs}(113)A$
  substrates}},\ }\href {https://doi.org/10.1103/PhysRevB.72.104430} {\bibfield
   {journal} {\bibinfo  {journal} {Phys. Rev. B}\ }\textbf {\bibinfo {volume}
  {72}},\ \bibinfo {pages} {104430} (\bibinfo {year} {2005})}\BibitemShut
  {NoStop}%
\bibitem [{\citenamefont {Limmer}\ \emph {et~al.}(2006)\citenamefont {Limmer},
  \citenamefont {Glunk}, \citenamefont {Daeubler}, \citenamefont {Hummel},
  \citenamefont {Schoch}, \citenamefont {Sauer}, \citenamefont {Bihler},
  \citenamefont {Huebl}, \citenamefont {Brandt},\ and\ \citenamefont
  {Goennenwein}}]{Limmer2006PRB}%
  \BibitemOpen
  \bibfield  {author} {\bibinfo {author} {\bibfnamefont {W.}~\bibnamefont
  {Limmer}}, \bibinfo {author} {\bibfnamefont {M.}~\bibnamefont {Glunk}},
  \bibinfo {author} {\bibfnamefont {J.}~\bibnamefont {Daeubler}}, \bibinfo
  {author} {\bibfnamefont {T.}~\bibnamefont {Hummel}}, \bibinfo {author}
  {\bibfnamefont {W.}~\bibnamefont {Schoch}}, \bibinfo {author} {\bibfnamefont
  {R.}~\bibnamefont {Sauer}}, \bibinfo {author} {\bibfnamefont
  {C.}~\bibnamefont {Bihler}}, \bibinfo {author} {\bibfnamefont
  {H.}~\bibnamefont {Huebl}}, \bibinfo {author} {\bibfnamefont {M.~S.}\
  \bibnamefont {Brandt}},\ and\ \bibinfo {author} {\bibfnamefont {S.~T.~B.}\
  \bibnamefont {Goennenwein}},\ }\bibfield  {title} {\bibinfo {title}
  {{Angle-dependent magnetotransport in cubic and tetragonal ferromagnets:
  Application to (001)- and $(113)A$-oriented
  $(\mathrm{Ga},\mathrm{Mn})\mathrm{As}$}},\ }\href
  {https://doi.org/10.1103/PhysRevB.74.205205} {\bibfield  {journal} {\bibinfo
  {journal} {Phys. Rev. B}\ }\textbf {\bibinfo {volume} {74}},\ \bibinfo
  {pages} {205205} (\bibinfo {year} {2006})}\BibitemShut {NoStop}%
\bibitem [{\citenamefont {Rushforth}\ \emph {et~al.}(2007)\citenamefont
  {Rushforth}, \citenamefont {V\'yborn\'y}, \citenamefont {King}, \citenamefont
  {Edmonds}, \citenamefont {Campion}, \citenamefont {Foxon}, \citenamefont
  {Wunderlich}, \citenamefont {Irvine}, \citenamefont
  {Va\ifmmode~\check{s}\else \v{s}\fi{}ek}, \citenamefont {Nov\'ak},
  \citenamefont {Olejn\'{\i}k}, \citenamefont {Sinova}, \citenamefont
  {Jungwirth},\ and\ \citenamefont {Gallagher}}]{Rushforth2007PRL}%
  \BibitemOpen
  \bibfield  {author} {\bibinfo {author} {\bibfnamefont {A.~W.}\ \bibnamefont
  {Rushforth}}, \bibinfo {author} {\bibfnamefont {K.}~\bibnamefont
  {V\'yborn\'y}}, \bibinfo {author} {\bibfnamefont {C.~S.}\ \bibnamefont
  {King}}, \bibinfo {author} {\bibfnamefont {K.~W.}\ \bibnamefont {Edmonds}},
  \bibinfo {author} {\bibfnamefont {R.~P.}\ \bibnamefont {Campion}}, \bibinfo
  {author} {\bibfnamefont {C.~T.}\ \bibnamefont {Foxon}}, \bibinfo {author}
  {\bibfnamefont {J.}~\bibnamefont {Wunderlich}}, \bibinfo {author}
  {\bibfnamefont {A.~C.}\ \bibnamefont {Irvine}}, \bibinfo {author}
  {\bibfnamefont {P.}~\bibnamefont {Va\ifmmode~\check{s}\else \v{s}\fi{}ek}},
  \bibinfo {author} {\bibfnamefont {V.}~\bibnamefont {Nov\'ak}}, \bibinfo
  {author} {\bibfnamefont {K.}~\bibnamefont {Olejn\'{\i}k}}, \bibinfo {author}
  {\bibfnamefont {J.}~\bibnamefont {Sinova}}, \bibinfo {author} {\bibfnamefont
  {T.}~\bibnamefont {Jungwirth}},\ and\ \bibinfo {author} {\bibfnamefont
  {B.~L.}\ \bibnamefont {Gallagher}},\ }\bibfield  {title} {\bibinfo {title}
  {{Anisotropic Magnetoresistance Components in (Ga,Mn)As}},\ }\href
  {https://doi.org/10.1103/PhysRevLett.99.147207} {\bibfield  {journal}
  {\bibinfo  {journal} {Phys. Rev. Lett.}\ }\textbf {\bibinfo {volume} {99}},\
  \bibinfo {pages} {147207} (\bibinfo {year} {2007})}\BibitemShut {NoStop}%
\bibitem [{\citenamefont {Bason}\ \emph {et~al.}(2009)\citenamefont {Bason},
  \citenamefont {Hoffman}, \citenamefont {Ahn},\ and\ \citenamefont
  {Klein}}]{Bason2009PRB}%
  \BibitemOpen
  \bibfield  {author} {\bibinfo {author} {\bibfnamefont {Y.}~\bibnamefont
  {Bason}}, \bibinfo {author} {\bibfnamefont {J.}~\bibnamefont {Hoffman}},
  \bibinfo {author} {\bibfnamefont {C.~H.}\ \bibnamefont {Ahn}},\ and\ \bibinfo
  {author} {\bibfnamefont {L.}~\bibnamefont {Klein}},\ }\bibfield  {title}
  {\bibinfo {title} {{Magnetoresistance tensor of
  ${\text{La}}_{0.8}{\text{Sr}}_{0.2}{\text{MnO}}_{3}$}},\ }\href
  {https://doi.org/10.1103/PhysRevB.79.092406} {\bibfield  {journal} {\bibinfo
  {journal} {Phys. Rev. B}\ }\textbf {\bibinfo {volume} {79}},\ \bibinfo
  {pages} {092406} (\bibinfo {year} {2009})}\BibitemShut {NoStop}%
\bibitem [{\citenamefont {Naftalis}\ \emph {et~al.}(2011)\citenamefont
  {Naftalis}, \citenamefont {Kaplan}, \citenamefont {Schultz}, \citenamefont
  {Vaz}, \citenamefont {Moyer}, \citenamefont {Ahn},\ and\ \citenamefont
  {Klein}}]{Naftalis2011PRB}%
  \BibitemOpen
  \bibfield  {author} {\bibinfo {author} {\bibfnamefont {N.}~\bibnamefont
  {Naftalis}}, \bibinfo {author} {\bibfnamefont {A.}~\bibnamefont {Kaplan}},
  \bibinfo {author} {\bibfnamefont {M.}~\bibnamefont {Schultz}}, \bibinfo
  {author} {\bibfnamefont {C.~A.~F.}\ \bibnamefont {Vaz}}, \bibinfo {author}
  {\bibfnamefont {J.~A.}\ \bibnamefont {Moyer}}, \bibinfo {author}
  {\bibfnamefont {C.~H.}\ \bibnamefont {Ahn}},\ and\ \bibinfo {author}
  {\bibfnamefont {L.}~\bibnamefont {Klein}},\ }\bibfield  {title} {\bibinfo
  {title} {{Field-dependent anisotropic magnetoresistance and planar Hall
  effect in epitaxial magnetite thin films}},\ }\href
  {https://doi.org/10.1103/PhysRevB.84.094441} {\bibfield  {journal} {\bibinfo
  {journal} {Phys. Rev. B}\ }\textbf {\bibinfo {volume} {84}},\ \bibinfo
  {pages} {094441} (\bibinfo {year} {2011})}\BibitemShut {NoStop}%
\bibitem [{\citenamefont {Annadi}\ \emph {et~al.}(2013)\citenamefont {Annadi},
  \citenamefont {Huang}, \citenamefont {Gopinadhan}, \citenamefont {Wang},
  \citenamefont {Srivastava}, \citenamefont {Liu}, \citenamefont {Ma},
  \citenamefont {Sarkar}, \citenamefont {Venkatesan},\ and\ \citenamefont
  {Ariando}}]{Annadi2013PRB}%
  \BibitemOpen
  \bibfield  {author} {\bibinfo {author} {\bibfnamefont {A.}~\bibnamefont
  {Annadi}}, \bibinfo {author} {\bibfnamefont {Z.}~\bibnamefont {Huang}},
  \bibinfo {author} {\bibfnamefont {K.}~\bibnamefont {Gopinadhan}}, \bibinfo
  {author} {\bibfnamefont {X.~R.}\ \bibnamefont {Wang}}, \bibinfo {author}
  {\bibfnamefont {A.}~\bibnamefont {Srivastava}}, \bibinfo {author}
  {\bibfnamefont {Z.~Q.}\ \bibnamefont {Liu}}, \bibinfo {author} {\bibfnamefont
  {H.~H.}\ \bibnamefont {Ma}}, \bibinfo {author} {\bibfnamefont {T.~P.}\
  \bibnamefont {Sarkar}}, \bibinfo {author} {\bibfnamefont {T.}~\bibnamefont
  {Venkatesan}},\ and\ \bibinfo {author} {\bibnamefont {Ariando}},\ }\bibfield
  {title} {\bibinfo {title} {{Fourfold oscillation in anisotropic
  magnetoresistance and planar Hall effect at the LaAlO${}_{3}$/SrTiO${}_{3}$
  heterointerfaces: Effect of carrier confinement and electric field on
  magnetic interactions}},\ }\href {https://doi.org/10.1103/PhysRevB.87.201102}
  {\bibfield  {journal} {\bibinfo  {journal} {Phys. Rev. B}\ }\textbf {\bibinfo
  {volume} {87}},\ \bibinfo {pages} {201102} (\bibinfo {year}
  {2013})}\BibitemShut {NoStop}%
\bibitem [{\citenamefont {Ding}\ \emph {et~al.}(2013)\citenamefont {Ding},
  \citenamefont {Li}, \citenamefont {Zhu}, \citenamefont {Ma}, \citenamefont
  {Won},\ and\ \citenamefont {Wu}}]{DingZ2013jap}%
  \BibitemOpen
  \bibfield  {author} {\bibinfo {author} {\bibfnamefont {Z.}~\bibnamefont
  {Ding}}, \bibinfo {author} {\bibfnamefont {J.~X.}\ \bibnamefont {Li}},
  \bibinfo {author} {\bibfnamefont {J.}~\bibnamefont {Zhu}}, \bibinfo {author}
  {\bibfnamefont {T.~P.}\ \bibnamefont {Ma}}, \bibinfo {author} {\bibfnamefont
  {C.}~\bibnamefont {Won}},\ and\ \bibinfo {author} {\bibfnamefont {Y.~Z.}\
  \bibnamefont {Wu}},\ }\bibfield  {title} {\bibinfo {title}
  {{Three-dimensional mapping of the anisotropic magnetoresistance in
  Fe${}_3$O${}_4$ single crystal thin films}},\ }\href
  {https://doi.org/10.1063/1.4796178} {\bibfield  {journal} {\bibinfo
  {journal} {Journal of Applied Physics}\ }\textbf {\bibinfo {volume} {113}},\
  \bibinfo {pages} {17B103} (\bibinfo {year} {2013})}\BibitemShut {NoStop}%
\bibitem [{\citenamefont {Xiao}\ \emph {et~al.}(2015)\citenamefont {Xiao},
  \citenamefont {Li}, \citenamefont {Ding},\ and\ \citenamefont
  {Wu}}]{xiao2015jap2}%
  \BibitemOpen
  \bibfield  {author} {\bibinfo {author} {\bibfnamefont {X.}~\bibnamefont
  {Xiao}}, \bibinfo {author} {\bibfnamefont {J.~X.}\ \bibnamefont {Li}},
  \bibinfo {author} {\bibfnamefont {Z.}~\bibnamefont {Ding}},\ and\ \bibinfo
  {author} {\bibfnamefont {Y.~Z.}\ \bibnamefont {Wu}},\ }\bibfield  {title}
  {\bibinfo {title} {{Four-fold symmetric anisotropic magnetoresistance of
  single-crystalline Ni(001) film}},\ }\href
  {https://doi.org/10.1063/1.4936175} {\bibfield  {journal} {\bibinfo
  {journal} {Journal of Applied Physics}\ }\textbf {\bibinfo {volume} {118}},\
  \bibinfo {pages} {203905} (\bibinfo {year} {2015})}\BibitemShut {NoStop}%
\bibitem [{\citenamefont {Oogane}\ \emph {et~al.}(2018)\citenamefont {Oogane},
  \citenamefont {McFadden}, \citenamefont {Kota}, \citenamefont {Brown-Heft},
  \citenamefont {Tsunoda}, \citenamefont {Ando},\ and\ \citenamefont
  {Palmstrøm}}]{Oogane2018jjap}%
  \BibitemOpen
  \bibfield  {author} {\bibinfo {author} {\bibfnamefont {M.}~\bibnamefont
  {Oogane}}, \bibinfo {author} {\bibfnamefont {A.~P.}\ \bibnamefont
  {McFadden}}, \bibinfo {author} {\bibfnamefont {Y.}~\bibnamefont {Kota}},
  \bibinfo {author} {\bibfnamefont {T.~L.}\ \bibnamefont {Brown-Heft}},
  \bibinfo {author} {\bibfnamefont {M.}~\bibnamefont {Tsunoda}}, \bibinfo
  {author} {\bibfnamefont {Y.}~\bibnamefont {Ando}},\ and\ \bibinfo {author}
  {\bibfnamefont {C.~J.}\ \bibnamefont {Palmstrøm}},\ }\bibfield  {title}
  {\bibinfo {title} {{Fourfold symmetric anisotropic magnetoresistance in
  half-metallic Co${}_2$MnSi Heusler alloy thin films}},\ }\href
  {https://doi.org/10.7567/JJAP.57.063001} {\bibfield  {journal} {\bibinfo
  {journal} {Japanese Journal of Applied Physics}\ }\textbf {\bibinfo {volume}
  {57}},\ \bibinfo {pages} {063001} (\bibinfo {year} {2018})}\BibitemShut
  {NoStop}%
\bibitem [{\citenamefont {Ahadi}\ \emph {et~al.}(2019)\citenamefont {Ahadi},
  \citenamefont {Lu}, \citenamefont {Salmani-Rezaie}, \citenamefont {Marshall},
  \citenamefont {Rondinelli},\ and\ \citenamefont {Stemmer}}]{Ahadi2019PRB}%
  \BibitemOpen
  \bibfield  {author} {\bibinfo {author} {\bibfnamefont {K.}~\bibnamefont
  {Ahadi}}, \bibinfo {author} {\bibfnamefont {X.}~\bibnamefont {Lu}}, \bibinfo
  {author} {\bibfnamefont {S.}~\bibnamefont {Salmani-Rezaie}}, \bibinfo
  {author} {\bibfnamefont {P.~B.}\ \bibnamefont {Marshall}}, \bibinfo {author}
  {\bibfnamefont {J.~M.}\ \bibnamefont {Rondinelli}},\ and\ \bibinfo {author}
  {\bibfnamefont {S.}~\bibnamefont {Stemmer}},\ }\bibfield  {title} {\bibinfo
  {title} {{Anisotropic magnetoresistance in the itinerant antiferromagnetic
  $\mathrm{EuTi}{\mathrm{O}}_{3}$}},\ }\href
  {https://doi.org/10.1103/PhysRevB.99.041106} {\bibfield  {journal} {\bibinfo
  {journal} {Phys. Rev. B}\ }\textbf {\bibinfo {volume} {99}},\ \bibinfo
  {pages} {041106} (\bibinfo {year} {2019})}\BibitemShut {NoStop}%
\bibitem [{\citenamefont {Zeng}\ \emph {et~al.}(2020)\citenamefont {Zeng},
  \citenamefont {Ren}, \citenamefont {Li}, \citenamefont {Zeng}, \citenamefont
  {Jia}, \citenamefont {Miao}, \citenamefont {Hoffmann}, \citenamefont {Zhang},
  \citenamefont {Wu},\ and\ \citenamefont {Yuan}}]{ZengFL2020PRL}%
  \BibitemOpen
  \bibfield  {author} {\bibinfo {author} {\bibfnamefont {F.~L.}\ \bibnamefont
  {Zeng}}, \bibinfo {author} {\bibfnamefont {Z.~Y.}\ \bibnamefont {Ren}},
  \bibinfo {author} {\bibfnamefont {Y.}~\bibnamefont {Li}}, \bibinfo {author}
  {\bibfnamefont {J.~Y.}\ \bibnamefont {Zeng}}, \bibinfo {author}
  {\bibfnamefont {M.~W.}\ \bibnamefont {Jia}}, \bibinfo {author} {\bibfnamefont
  {J.}~\bibnamefont {Miao}}, \bibinfo {author} {\bibfnamefont {A.}~\bibnamefont
  {Hoffmann}}, \bibinfo {author} {\bibfnamefont {W.}~\bibnamefont {Zhang}},
  \bibinfo {author} {\bibfnamefont {Y.~Z.}\ \bibnamefont {Wu}},\ and\ \bibinfo
  {author} {\bibfnamefont {Z.}~\bibnamefont {Yuan}},\ }\bibfield  {title}
  {\bibinfo {title} {{Intrinsic Mechanism for Anisotropic Magnetoresistance and
  Experimental Confirmation in
  ${\mathrm{Co}}_{x}{\mathrm{Fe}}_{1\ensuremath{-}x}$ Single-Crystal Films}},\
  }\href {https://doi.org/10.1103/PhysRevLett.125.097201} {\bibfield  {journal}
  {\bibinfo  {journal} {Phys. Rev. Lett.}\ }\textbf {\bibinfo {volume} {125}},\
  \bibinfo {pages} {097201} (\bibinfo {year} {2020})}\BibitemShut {NoStop}%
\bibitem [{\citenamefont {Wadehra}\ \emph {et~al.}(2020)\citenamefont
  {Wadehra}, \citenamefont {Tomar}, \citenamefont {Varma}, \citenamefont
  {Gopal}, \citenamefont {Singh}, \citenamefont {Dattagupta},\ and\
  \citenamefont {Chakraverty}}]{Wadehra2020NC}%
  \BibitemOpen
  \bibfield  {author} {\bibinfo {author} {\bibfnamefont {N.}~\bibnamefont
  {Wadehra}}, \bibinfo {author} {\bibfnamefont {R.}~\bibnamefont {Tomar}},
  \bibinfo {author} {\bibfnamefont {R.~M.}\ \bibnamefont {Varma}}, \bibinfo
  {author} {\bibfnamefont {R.~K.}\ \bibnamefont {Gopal}}, \bibinfo {author}
  {\bibfnamefont {Y.}~\bibnamefont {Singh}}, \bibinfo {author} {\bibfnamefont
  {S.}~\bibnamefont {Dattagupta}},\ and\ \bibinfo {author} {\bibfnamefont
  {S.}~\bibnamefont {Chakraverty}},\ }\bibfield  {title} {\bibinfo {title}
  {{Planar Hall effect and anisotropic magnetoresistance in polar-polar
  interface of LaVO${}_{3}$-KTaO${}_{3}$ with strong spin-orbit coupling}},\
  }\href {https://doi.org/10.1038/s41467-020-14689-z} {\bibfield  {journal}
  {\bibinfo  {journal} {Nature Communications}\ }\textbf {\bibinfo {volume}
  {11}},\ \bibinfo {pages} {874} (\bibinfo {year} {2020})}\BibitemShut
  {NoStop}%
\bibitem [{\citenamefont {Battilomo}\ \emph {et~al.}(2021)\citenamefont
  {Battilomo}, \citenamefont {Scopigno},\ and\ \citenamefont
  {Ortix}}]{Battilomo2021PRB}%
  \BibitemOpen
  \bibfield  {author} {\bibinfo {author} {\bibfnamefont {R.}~\bibnamefont
  {Battilomo}}, \bibinfo {author} {\bibfnamefont {N.}~\bibnamefont
  {Scopigno}},\ and\ \bibinfo {author} {\bibfnamefont {C.}~\bibnamefont
  {Ortix}},\ }\bibfield  {title} {\bibinfo {title} {{Anomalous planar Hall
  effect in two-dimensional trigonal crystals}},\ }\href
  {https://doi.org/10.1103/PhysRevResearch.3.L012006} {\bibfield  {journal}
  {\bibinfo  {journal} {Phys. Rev. Research}\ }\textbf {\bibinfo {volume}
  {3}},\ \bibinfo {pages} {L012006} (\bibinfo {year} {2021})}\BibitemShut
  {NoStop}%
\bibitem [{\citenamefont {Dai}\ \emph {et~al.}(2022)\citenamefont {Dai},
  \citenamefont {Zhao}, \citenamefont {Ma}, \citenamefont {Tang}, \citenamefont
  {Qiu}, \citenamefont {Liu}, \citenamefont {Yuan},\ and\ \citenamefont
  {Zhou}}]{DaiY2022PRL}%
  \BibitemOpen
  \bibfield  {author} {\bibinfo {author} {\bibfnamefont {Y.}~\bibnamefont
  {Dai}}, \bibinfo {author} {\bibfnamefont {Y.~W.}\ \bibnamefont {Zhao}},
  \bibinfo {author} {\bibfnamefont {L.}~\bibnamefont {Ma}}, \bibinfo {author}
  {\bibfnamefont {M.}~\bibnamefont {Tang}}, \bibinfo {author} {\bibfnamefont
  {X.~P.}\ \bibnamefont {Qiu}}, \bibinfo {author} {\bibfnamefont
  {Y.}~\bibnamefont {Liu}}, \bibinfo {author} {\bibfnamefont {Z.}~\bibnamefont
  {Yuan}},\ and\ \bibinfo {author} {\bibfnamefont {S.~M.}\ \bibnamefont
  {Zhou}},\ }\bibfield  {title} {\bibinfo {title} {{Fourfold Anisotropic
  Magnetoresistance of ${\mathrm{L}1}_{0}$ FePt Due to Relaxation Time
  Anisotropy}},\ }\href {https://doi.org/10.1103/PhysRevLett.128.247202}
  {\bibfield  {journal} {\bibinfo  {journal} {Phys. Rev. Lett.}\ }\textbf
  {\bibinfo {volume} {128}},\ \bibinfo {pages} {247202} (\bibinfo {year}
  {2022})}\BibitemShut {NoStop}%
\bibitem [{\citenamefont {Song}\ \emph {et~al.}(2022)\citenamefont {Song},
  \citenamefont {Oh}, \citenamefont {Ko}, \citenamefont {Lee}, \citenamefont
  {Kim}, \citenamefont {Zhu}, \citenamefont {Yang}, \citenamefont {Li},\ and\
  \citenamefont {Noh}}]{Song2022NC}%
  \BibitemOpen
  \bibfield  {author} {\bibinfo {author} {\bibfnamefont {J.}~\bibnamefont
  {Song}}, \bibinfo {author} {\bibfnamefont {T.}~\bibnamefont {Oh}}, \bibinfo
  {author} {\bibfnamefont {E.~K.}\ \bibnamefont {Ko}}, \bibinfo {author}
  {\bibfnamefont {J.~H.}\ \bibnamefont {Lee}}, \bibinfo {author} {\bibfnamefont
  {W.~J.}\ \bibnamefont {Kim}}, \bibinfo {author} {\bibfnamefont
  {Y.}~\bibnamefont {Zhu}}, \bibinfo {author} {\bibfnamefont {B.-J.}\
  \bibnamefont {Yang}}, \bibinfo {author} {\bibfnamefont {Y.}~\bibnamefont
  {Li}},\ and\ \bibinfo {author} {\bibfnamefont {T.~W.}\ \bibnamefont {Noh}},\
  }\bibfield  {title} {\bibinfo {title} {{Higher harmonics in planar Hall
  effect induced by cluster magnetic multipoles}},\ }\href
  {https://doi.org/10.1038/s41467-022-34189-6} {\bibfield  {journal} {\bibinfo
  {journal} {Nature Communications}\ }\textbf {\bibinfo {volume} {13}},\
  \bibinfo {pages} {6501} (\bibinfo {year} {2022})}\BibitemShut {NoStop}%
\bibitem [{\citenamefont {Yahagi}\ \emph {et~al.}(2020)\citenamefont {Yahagi},
  \citenamefont {Miura},\ and\ \citenamefont {Sakuma}}]{Yahagi2020jpsj}%
  \BibitemOpen
  \bibfield  {author} {\bibinfo {author} {\bibfnamefont {Y.}~\bibnamefont
  {Yahagi}}, \bibinfo {author} {\bibfnamefont {D.}~\bibnamefont {Miura}},\ and\
  \bibinfo {author} {\bibfnamefont {A.}~\bibnamefont {Sakuma}},\ }\bibfield
  {title} {\bibinfo {title} {{Theoretical Study on Four-fold Symmetric
  Anisotropic Magnetoresistance Effect in Cubic Single-crystal Ferromagnetic
  Model}},\ }\href {https://doi.org/10.7566/JPSJ.89.044714} {\bibfield
  {journal} {\bibinfo  {journal} {Journal of the Physical Society of Japan}\
  }\textbf {\bibinfo {volume} {89}},\ \bibinfo {pages} {044714} (\bibinfo
  {year} {2020})}\BibitemShut {NoStop}%
\bibitem [{\citenamefont {Zawadzki}(2017)}]{Zawadzki2017jpcm}%
  \BibitemOpen
  \bibfield  {author} {\bibinfo {author} {\bibfnamefont {W.}~\bibnamefont
  {Zawadzki}},\ }\bibfield  {title} {\bibinfo {title} {{Semirelativity in
  semiconductors: a review}},\ }\href
  {https://doi.org/10.1088/1361-648X/aa7932} {\bibfield  {journal} {\bibinfo
  {journal} {Journal of Physics: Condensed Matter}\ }\textbf {\bibinfo {volume}
  {29}},\ \bibinfo {pages} {373004} (\bibinfo {year} {2017})}\BibitemShut
  {NoStop}%
\bibitem [{\citenamefont {Andreev}\ and\ \citenamefont
  {Spivak}(2018)}]{Andreev2018PRL}%
  \BibitemOpen
  \bibfield  {author} {\bibinfo {author} {\bibfnamefont {A.~V.}\ \bibnamefont
  {Andreev}}\ and\ \bibinfo {author} {\bibfnamefont {B.~Z.}\ \bibnamefont
  {Spivak}},\ }\bibfield  {title} {\bibinfo {title} {{Longitudinal Negative
  Magnetoresistance and Magnetotransport Phenomena in Conventional and
  Topological Conductors}},\ }\href
  {https://doi.org/10.1103/PhysRevLett.120.026601} {\bibfield  {journal}
  {\bibinfo  {journal} {Phys. Rev. Lett.}\ }\textbf {\bibinfo {volume} {120}},\
  \bibinfo {pages} {026601} (\bibinfo {year} {2018})}\BibitemShut {NoStop}%
\bibitem [{\citenamefont {Hasan}\ and\ \citenamefont
  {Kane}(2010)}]{Hasan2010RMP}%
  \BibitemOpen
  \bibfield  {author} {\bibinfo {author} {\bibfnamefont {M.~Z.}\ \bibnamefont
  {Hasan}}\ and\ \bibinfo {author} {\bibfnamefont {C.~L.}\ \bibnamefont
  {Kane}},\ }\bibfield  {title} {\bibinfo {title} {{\textit{Colloquium} :
  Topological insulators}},\ }\href
  {https://doi.org/10.1103/RevModPhys.82.3045} {\bibfield  {journal} {\bibinfo
  {journal} {Rev. Mod. Phys.}\ }\textbf {\bibinfo {volume} {82}},\ \bibinfo
  {pages} {3045} (\bibinfo {year} {2010})}\BibitemShut {NoStop}%
\bibitem [{\citenamefont {Qi}\ and\ \citenamefont {Zhang}(2011)}]{QiZhangRMP}%
  \BibitemOpen
  \bibfield  {author} {\bibinfo {author} {\bibfnamefont {X.-L.}\ \bibnamefont
  {Qi}}\ and\ \bibinfo {author} {\bibfnamefont {S.-C.}\ \bibnamefont {Zhang}},\
  }\bibfield  {title} {\bibinfo {title} {{Topological insulators and
  superconductors}},\ }\href {https://doi.org/10.1103/RevModPhys.83.1057}
  {\bibfield  {journal} {\bibinfo  {journal} {Rev. Mod. Phys.}\ }\textbf
  {\bibinfo {volume} {83}},\ \bibinfo {pages} {1057} (\bibinfo {year}
  {2011})}\BibitemShut {NoStop}%
\bibitem [{\citenamefont {Weng}\ \emph {et~al.}(2015)\citenamefont {Weng},
  \citenamefont {Yu}, \citenamefont {Hu}, \citenamefont {Dai},\ and\
  \citenamefont {Fang}}]{WengQAHE2015}%
  \BibitemOpen
  \bibfield  {author} {\bibinfo {author} {\bibfnamefont {H.}~\bibnamefont
  {Weng}}, \bibinfo {author} {\bibfnamefont {R.}~\bibnamefont {Yu}}, \bibinfo
  {author} {\bibfnamefont {X.}~\bibnamefont {Hu}}, \bibinfo {author}
  {\bibfnamefont {X.}~\bibnamefont {Dai}},\ and\ \bibinfo {author}
  {\bibfnamefont {Z.}~\bibnamefont {Fang}},\ }\bibfield  {title} {\bibinfo
  {title} {{Quantum anomalous Hall effect and related topological electronic
  states}},\ }\href {https://doi.org/10.1080/00018732.2015.1068524} {\bibfield
  {journal} {\bibinfo  {journal} {Advances in Physics}\ }\textbf {\bibinfo
  {volume} {64}},\ \bibinfo {pages} {227} (\bibinfo {year} {2015})}\BibitemShut
  {NoStop}%
\bibitem [{\citenamefont {Chang}\ \emph {et~al.}(2023)\citenamefont {Chang},
  \citenamefont {Liu},\ and\ \citenamefont {MacDonald}}]{Chang2023RMP}%
  \BibitemOpen
  \bibfield  {author} {\bibinfo {author} {\bibfnamefont {C.-Z.}\ \bibnamefont
  {Chang}}, \bibinfo {author} {\bibfnamefont {C.-X.}\ \bibnamefont {Liu}},\
  and\ \bibinfo {author} {\bibfnamefont {A.~H.}\ \bibnamefont {MacDonald}},\
  }\bibfield  {title} {\bibinfo {title} {{Colloquium: Quantum anomalous Hall
  effect}},\ }\href {https://doi.org/10.1103/RevModPhys.95.011002} {\bibfield
  {journal} {\bibinfo  {journal} {Rev. Mod. Phys.}\ }\textbf {\bibinfo {volume}
  {95}},\ \bibinfo {pages} {011002} (\bibinfo {year} {2023})}\BibitemShut
  {NoStop}%
\bibitem [{\citenamefont {Xiao}\ \emph {et~al.}(2010)\citenamefont {Xiao},
  \citenamefont {Chang},\ and\ \citenamefont {Niu}}]{Xiao2010RMP}%
  \BibitemOpen
  \bibfield  {author} {\bibinfo {author} {\bibfnamefont {D.}~\bibnamefont
  {Xiao}}, \bibinfo {author} {\bibfnamefont {M.-C.}\ \bibnamefont {Chang}},\
  and\ \bibinfo {author} {\bibfnamefont {Q.}~\bibnamefont {Niu}},\ }\bibfield
  {title} {\bibinfo {title} {{Berry phase effects on electronic properties}},\
  }\href {https://doi.org/10.1103/RevModPhys.82.1959} {\bibfield  {journal}
  {\bibinfo  {journal} {Rev. Mod. Phys.}\ }\textbf {\bibinfo {volume} {82}},\
  \bibinfo {pages} {1959} (\bibinfo {year} {2010})}\BibitemShut {NoStop}%
\bibitem [{\citenamefont {Liu}\ and\ \citenamefont {Dai}(2021)}]{Liu2021NRP}%
  \BibitemOpen
  \bibfield  {author} {\bibinfo {author} {\bibfnamefont {J.}~\bibnamefont
  {Liu}}\ and\ \bibinfo {author} {\bibfnamefont {X.}~\bibnamefont {Dai}},\
  }\bibfield  {title} {\bibinfo {title} {{Orbital magnetic states in moir{\'e}
  graphene systems}},\ }\href {https://doi.org/10.1038/s42254-021-00297-3}
  {\bibfield  {journal} {\bibinfo  {journal} {Nature Reviews Physics}\ }\textbf
  {\bibinfo {volume} {3}},\ \bibinfo {pages} {367} (\bibinfo {year}
  {2021})}\BibitemShut {NoStop}%
\bibitem [{\citenamefont {Atencia}\ \emph {et~al.}(2024)\citenamefont
  {Atencia}, \citenamefont {Agarwal},\ and\ \citenamefont
  {Culcer}}]{atencia2024omm}%
  \BibitemOpen
  \bibfield  {author} {\bibinfo {author} {\bibfnamefont {R.~B.}\ \bibnamefont
  {Atencia}}, \bibinfo {author} {\bibfnamefont {A.}~\bibnamefont {Agarwal}},\
  and\ \bibinfo {author} {\bibfnamefont {D.}~\bibnamefont {Culcer}},\
  }\bibfield  {title} {\bibinfo {title} {Orbital angular momentum of bloch
  electrons: equilibrium formulation, magneto-electric phenomena, and the
  orbital hall effect},\ }\href {https://doi.org/10.1080/23746149.2024.2371972}
  {\bibfield  {journal} {\bibinfo  {journal} {Advances in Physics: X}\ }\textbf
  {\bibinfo {volume} {9}},\ \bibinfo {pages} {2371972} (\bibinfo {year}
  {2024})}\BibitemShut {NoStop}%
\bibitem [{\citenamefont {Yang}\ \emph {et~al.}(2020)\citenamefont {Yang},
  \citenamefont {Chang},\ and\ \citenamefont {Parkin}}]{YangSY2020PRR}%
  \BibitemOpen
  \bibfield  {author} {\bibinfo {author} {\bibfnamefont {S.-Y.}\ \bibnamefont
  {Yang}}, \bibinfo {author} {\bibfnamefont {K.}~\bibnamefont {Chang}},\ and\
  \bibinfo {author} {\bibfnamefont {S.~S.~P.}\ \bibnamefont {Parkin}},\
  }\bibfield  {title} {\bibinfo {title} {{Large planar Hall effect in bismuth
  thin films}},\ }\href {https://doi.org/10.1103/PhysRevResearch.2.022029}
  {\bibfield  {journal} {\bibinfo  {journal} {Phys. Rev. Research}\ }\textbf
  {\bibinfo {volume} {2}},\ \bibinfo {pages} {022029} (\bibinfo {year}
  {2020})}\BibitemShut {NoStop}%
\bibitem [{\citenamefont {Wu}\ \emph {et~al.}(2022)\citenamefont {Wu},
  \citenamefont {Tu}, \citenamefont {Nie}, \citenamefont {Miao}, \citenamefont
  {Gao}, \citenamefont {Han}, \citenamefont {Zhu}, \citenamefont {Zhou},
  \citenamefont {Ning},\ and\ \citenamefont {Tian}}]{Wu2022NL}%
  \BibitemOpen
  \bibfield  {author} {\bibinfo {author} {\bibfnamefont {M.}~\bibnamefont
  {Wu}}, \bibinfo {author} {\bibfnamefont {D.}~\bibnamefont {Tu}}, \bibinfo
  {author} {\bibfnamefont {Y.}~\bibnamefont {Nie}}, \bibinfo {author}
  {\bibfnamefont {S.}~\bibnamefont {Miao}}, \bibinfo {author} {\bibfnamefont
  {W.}~\bibnamefont {Gao}}, \bibinfo {author} {\bibfnamefont {Y.}~\bibnamefont
  {Han}}, \bibinfo {author} {\bibfnamefont {X.}~\bibnamefont {Zhu}}, \bibinfo
  {author} {\bibfnamefont {J.}~\bibnamefont {Zhou}}, \bibinfo {author}
  {\bibfnamefont {W.}~\bibnamefont {Ning}},\ and\ \bibinfo {author}
  {\bibfnamefont {M.}~\bibnamefont {Tian}},\ }\bibfield  {title} {\bibinfo
  {title} {{Novel $\pi$/2-Periodic Planar Hall Effect Due to Orbital Magnetic
  Moments in MnBi${}_2$Te${}_4$}},\ }\href
  {https://doi.org/10.1021/acs.nanolett.1c03232} {\bibfield  {journal}
  {\bibinfo  {journal} {Nano Letters}\ }\textbf {\bibinfo {volume} {22}},\
  \bibinfo {pages} {73} (\bibinfo {year} {2022})}\BibitemShut {NoStop}%
\bibitem [{\citenamefont {Yamada}\ and\ \citenamefont
  {Fuseya}(2021)}]{Yamada2021PRB}%
  \BibitemOpen
  \bibfield  {author} {\bibinfo {author} {\bibfnamefont {A.}~\bibnamefont
  {Yamada}}\ and\ \bibinfo {author} {\bibfnamefont {Y.}~\bibnamefont
  {Fuseya}},\ }\bibfield  {title} {\bibinfo {title} {{Angular dependence of
  magnetoresistance and planar Hall effect in semimetals in strong magnetic
  fields}},\ }\href {https://doi.org/10.1103/PhysRevB.103.125148} {\bibfield
  {journal} {\bibinfo  {journal} {Phys. Rev. B}\ }\textbf {\bibinfo {volume}
  {103}},\ \bibinfo {pages} {125148} (\bibinfo {year} {2021})}\BibitemShut
  {NoStop}%
\bibitem [{\citenamefont {Sundaram}\ and\ \citenamefont
  {Niu}(1999)}]{Sundaram1999PRB}%
  \BibitemOpen
  \bibfield  {author} {\bibinfo {author} {\bibfnamefont {G.}~\bibnamefont
  {Sundaram}}\ and\ \bibinfo {author} {\bibfnamefont {Q.}~\bibnamefont {Niu}},\
  }\bibfield  {title} {\bibinfo {title} {{Wave-packet dynamics in slowly
  perturbed crystals: Gradient corrections and Berry-phase effects}},\ }\href
  {https://doi.org/10.1103/PhysRevB.59.14915} {\bibfield  {journal} {\bibinfo
  {journal} {Phys. Rev. B}\ }\textbf {\bibinfo {volume} {59}},\ \bibinfo
  {pages} {14915} (\bibinfo {year} {1999})}\BibitemShut {NoStop}%
\bibitem [{\citenamefont {Xiao}\ \emph {et~al.}(2005)\citenamefont {Xiao},
  \citenamefont {Shi},\ and\ \citenamefont {Niu}}]{Xiao2005PRL}%
  \BibitemOpen
  \bibfield  {author} {\bibinfo {author} {\bibfnamefont {D.}~\bibnamefont
  {Xiao}}, \bibinfo {author} {\bibfnamefont {J.}~\bibnamefont {Shi}},\ and\
  \bibinfo {author} {\bibfnamefont {Q.}~\bibnamefont {Niu}},\ }\bibfield
  {title} {\bibinfo {title} {{Berry Phase Correction to Electron Density of
  States in Solids}},\ }\href {https://doi.org/10.1103/PhysRevLett.95.137204}
  {\bibfield  {journal} {\bibinfo  {journal} {Phys. Rev. Lett.}\ }\textbf
  {\bibinfo {volume} {95}},\ \bibinfo {pages} {137204} (\bibinfo {year}
  {2005})}\BibitemShut {NoStop}%
\bibitem [{\citenamefont {Thonhauser}\ \emph {et~al.}(2005)\citenamefont
  {Thonhauser}, \citenamefont {Ceresoli}, \citenamefont {Vanderbilt},\ and\
  \citenamefont {Resta}}]{Thonhauser2005PRL}%
  \BibitemOpen
  \bibfield  {author} {\bibinfo {author} {\bibfnamefont {T.}~\bibnamefont
  {Thonhauser}}, \bibinfo {author} {\bibfnamefont {D.}~\bibnamefont
  {Ceresoli}}, \bibinfo {author} {\bibfnamefont {D.}~\bibnamefont
  {Vanderbilt}},\ and\ \bibinfo {author} {\bibfnamefont {R.}~\bibnamefont
  {Resta}},\ }\bibfield  {title} {\bibinfo {title} {{Orbital Magnetization in
  Periodic Insulators}},\ }\href
  {https://doi.org/10.1103/PhysRevLett.95.137205} {\bibfield  {journal}
  {\bibinfo  {journal} {Phys. Rev. Lett.}\ }\textbf {\bibinfo {volume} {95}},\
  \bibinfo {pages} {137205} (\bibinfo {year} {2005})}\BibitemShut {NoStop}%
\bibitem [{\citenamefont {Ziman}(2001)}]{EPhonon2001Ziman}%
  \BibitemOpen
  \bibfield  {author} {\bibinfo {author} {\bibfnamefont {J.}~\bibnamefont
  {Ziman}},\ }\href {https://doi.org/10.1093/acprof:oso/9780198507796.001.0001}
  {\emph {\bibinfo {title} {{Electrons and Phonons: The Theory of Transport
  Phenomena in Solids}}}}\ (\bibinfo  {publisher} {Oxford University Press, New
  York},\ \bibinfo {year} {2001})\BibitemShut {NoStop}%
\bibitem [{\citenamefont {Dai}\ \emph {et~al.}(2017)\citenamefont {Dai},
  \citenamefont {Du},\ and\ \citenamefont {Lu}}]{DaiX2017PRL}%
  \BibitemOpen
  \bibfield  {author} {\bibinfo {author} {\bibfnamefont {X.}~\bibnamefont
  {Dai}}, \bibinfo {author} {\bibfnamefont {Z.~Z.}\ \bibnamefont {Du}},\ and\
  \bibinfo {author} {\bibfnamefont {H.-Z.}\ \bibnamefont {Lu}},\ }\bibfield
  {title} {\bibinfo {title} {{Negative Magnetoresistance without Chiral Anomaly
  in Topological Insulators}},\ }\href
  {https://doi.org/10.1103/PhysRevLett.119.166601} {\bibfield  {journal}
  {\bibinfo  {journal} {Phys. Rev. Lett.}\ }\textbf {\bibinfo {volume} {119}},\
  \bibinfo {pages} {166601} (\bibinfo {year} {2017})}\BibitemShut {NoStop}%
\bibitem [{\citenamefont {Nandy}\ \emph {et~al.}(2018)\citenamefont {Nandy},
  \citenamefont {Taraphder},\ and\ \citenamefont {Tewari}}]{Nandy2018SR}%
  \BibitemOpen
  \bibfield  {author} {\bibinfo {author} {\bibfnamefont {S.}~\bibnamefont
  {Nandy}}, \bibinfo {author} {\bibfnamefont {A.}~\bibnamefont {Taraphder}},\
  and\ \bibinfo {author} {\bibfnamefont {S.}~\bibnamefont {Tewari}},\
  }\bibfield  {title} {\bibinfo {title} {{Berry phase theory of planar Hall
  effect in topological insulators}},\ }\href
  {https://doi.org/10.1038/s41598-018-33258-5} {\bibfield  {journal} {\bibinfo
  {journal} {Scientific Reports}\ }\textbf {\bibinfo {volume} {8}},\ \bibinfo
  {pages} {14983} (\bibinfo {year} {2018})}\BibitemShut {NoStop}%
\bibitem [{\citenamefont {Gao}\ \emph {et~al.}(2017)\citenamefont {Gao},
  \citenamefont {Yang},\ and\ \citenamefont {Niu}}]{GaoY2017PRB}%
  \BibitemOpen
  \bibfield  {author} {\bibinfo {author} {\bibfnamefont {Y.}~\bibnamefont
  {Gao}}, \bibinfo {author} {\bibfnamefont {S.~A.}\ \bibnamefont {Yang}},\ and\
  \bibinfo {author} {\bibfnamefont {Q.}~\bibnamefont {Niu}},\ }\bibfield
  {title} {\bibinfo {title} {{Intrinsic relative magnetoconductivity of
  nonmagnetic metals}},\ }\href {https://doi.org/10.1103/PhysRevB.95.165135}
  {\bibfield  {journal} {\bibinfo  {journal} {Phys. Rev. B}\ }\textbf {\bibinfo
  {volume} {95}},\ \bibinfo {pages} {165135} (\bibinfo {year}
  {2017})}\BibitemShut {NoStop}%
\bibitem [{\citenamefont {Gao}(2019)}]{Gao2019FoP}%
  \BibitemOpen
  \bibfield  {author} {\bibinfo {author} {\bibfnamefont {Y.}~\bibnamefont
  {Gao}},\ }\bibfield  {title} {\bibinfo {title} {{Semiclassical dynamics and
  nonlinear charge current}},\ }\href
  {https://doi.org/10.1007/s11467-019-0887-2} {\bibfield  {journal} {\bibinfo
  {journal} {Frontiers of Physics}\ }\textbf {\bibinfo {volume} {14}},\
  \bibinfo {pages} {33404} (\bibinfo {year} {2019})}\BibitemShut {NoStop}%
\bibitem [{\citenamefont {Huang}\ \emph {et~al.}(2023)\citenamefont {Huang},
  \citenamefont {Feng}, \citenamefont {Wang}, \citenamefont {Xiao},\ and\
  \citenamefont {Yang}}]{HuangYX2023PRL}%
  \BibitemOpen
  \bibfield  {author} {\bibinfo {author} {\bibfnamefont {Y.-X.}\ \bibnamefont
  {Huang}}, \bibinfo {author} {\bibfnamefont {X.}~\bibnamefont {Feng}},
  \bibinfo {author} {\bibfnamefont {H.}~\bibnamefont {Wang}}, \bibinfo {author}
  {\bibfnamefont {C.}~\bibnamefont {Xiao}},\ and\ \bibinfo {author}
  {\bibfnamefont {S.~A.}\ \bibnamefont {Yang}},\ }\bibfield  {title} {\bibinfo
  {title} {{Intrinsic Nonlinear Planar Hall Effect}},\ }\href
  {https://doi.org/10.1103/PhysRevLett.130.126303} {\bibfield  {journal}
  {\bibinfo  {journal} {Phys. Rev. Lett.}\ }\textbf {\bibinfo {volume} {130}},\
  \bibinfo {pages} {126303} (\bibinfo {year} {2023})}\BibitemShut {NoStop}%
\bibitem [{\citenamefont {Shen}(2013)}]{shen2013TIDirac}%
  \BibitemOpen
  \bibfield  {author} {\bibinfo {author} {\bibfnamefont {S.-Q.}\ \bibnamefont
  {Shen}},\ }\href@noop {} {\emph {\bibinfo {title} {{Topological Insulators:
  Dirac Equation in Condensed Matters}}}}\ (\bibinfo  {publisher}
  {Springer-Verlag, Berlin},\ \bibinfo {year} {2013})\BibitemShut {NoStop}%
\bibitem [{\citenamefont {Armitage}\ \emph {et~al.}(2018)\citenamefont
  {Armitage}, \citenamefont {Mele},\ and\ \citenamefont
  {Vishwanath}}]{Armitage2018RMP}%
  \BibitemOpen
  \bibfield  {author} {\bibinfo {author} {\bibfnamefont {N.~P.}\ \bibnamefont
  {Armitage}}, \bibinfo {author} {\bibfnamefont {E.~J.}\ \bibnamefont {Mele}},\
  and\ \bibinfo {author} {\bibfnamefont {A.}~\bibnamefont {Vishwanath}},\
  }\bibfield  {title} {\bibinfo {title} {{Weyl and Dirac semimetals in
  three-dimensional solids}},\ }\href
  {https://doi.org/10.1103/RevModPhys.90.015001} {\bibfield  {journal}
  {\bibinfo  {journal} {Rev. Mod. Phys.}\ }\textbf {\bibinfo {volume} {90}},\
  \bibinfo {pages} {015001} (\bibinfo {year} {2018})}\BibitemShut {NoStop}%
\bibitem [{\citenamefont {Lv}\ \emph {et~al.}(2021)\citenamefont {Lv},
  \citenamefont {Qian},\ and\ \citenamefont {Ding}}]{Lv2021RMP}%
  \BibitemOpen
  \bibfield  {author} {\bibinfo {author} {\bibfnamefont {B.~Q.}\ \bibnamefont
  {Lv}}, \bibinfo {author} {\bibfnamefont {T.}~\bibnamefont {Qian}},\ and\
  \bibinfo {author} {\bibfnamefont {H.}~\bibnamefont {Ding}},\ }\bibfield
  {title} {\bibinfo {title} {{Experimental perspective on three-dimensional
  topological semimetals}},\ }\href
  {https://doi.org/10.1103/RevModPhys.93.025002} {\bibfield  {journal}
  {\bibinfo  {journal} {Rev. Mod. Phys.}\ }\textbf {\bibinfo {volume} {93}},\
  \bibinfo {pages} {025002} (\bibinfo {year} {2021})}\BibitemShut {NoStop}%
\bibitem [{\citenamefont {Guo}\ \emph {et~al.}(2010)\citenamefont {Guo},
  \citenamefont {Rosenberg}, \citenamefont {Refael},\ and\ \citenamefont
  {Franz}}]{GuoHM2010PRL}%
  \BibitemOpen
  \bibfield  {author} {\bibinfo {author} {\bibfnamefont {H.-M.}\ \bibnamefont
  {Guo}}, \bibinfo {author} {\bibfnamefont {G.}~\bibnamefont {Rosenberg}},
  \bibinfo {author} {\bibfnamefont {G.}~\bibnamefont {Refael}},\ and\ \bibinfo
  {author} {\bibfnamefont {M.}~\bibnamefont {Franz}},\ }\bibfield  {title}
  {\bibinfo {title} {{Topological Anderson Insulator in Three Dimensions}},\
  }\href {https://doi.org/10.1103/PhysRevLett.105.216601} {\bibfield  {journal}
  {\bibinfo  {journal} {Phys. Rev. Lett.}\ }\textbf {\bibinfo {volume} {105}},\
  \bibinfo {pages} {216601} (\bibinfo {year} {2010})}\BibitemShut {NoStop}%
\bibitem [{\citenamefont {Fu}\ \emph {et~al.}(2020)\citenamefont {Fu},
  \citenamefont {Wang},\ and\ \citenamefont {Shen}}]{FuBo2020PRL}%
  \BibitemOpen
  \bibfield  {author} {\bibinfo {author} {\bibfnamefont {B.}~\bibnamefont
  {Fu}}, \bibinfo {author} {\bibfnamefont {H.-W.}\ \bibnamefont {Wang}},\ and\
  \bibinfo {author} {\bibfnamefont {S.-Q.}\ \bibnamefont {Shen}},\ }\bibfield
  {title} {\bibinfo {title} {{Dirac Polarons and Resistivity Anomaly in
  ${\mathrm{ZrTe}}_{5}$ and ${\mathrm{HfTe}}_{5}$}},\ }\href
  {https://doi.org/10.1103/PhysRevLett.125.256601} {\bibfield  {journal}
  {\bibinfo  {journal} {Phys. Rev. Lett.}\ }\textbf {\bibinfo {volume} {125}},\
  \bibinfo {pages} {256601} (\bibinfo {year} {2020})}\BibitemShut {NoStop}%
\bibitem [{\citenamefont {Chang}\ and\ \citenamefont
  {Niu}(2008)}]{Chang2008jpcm}%
  \BibitemOpen
  \bibfield  {author} {\bibinfo {author} {\bibfnamefont {M.-C.}\ \bibnamefont
  {Chang}}\ and\ \bibinfo {author} {\bibfnamefont {Q.}~\bibnamefont {Niu}},\
  }\bibfield  {title} {\bibinfo {title} {{Berry curvature, orbital moment, and
  effective quantum theory of electrons in electromagnetic fields}},\ }\href
  {https://doi.org/10.1088/0953-8984/20/19/193202} {\bibfield  {journal}
  {\bibinfo  {journal} {Journal of Physics: Condensed Matter}\ }\textbf
  {\bibinfo {volume} {20}},\ \bibinfo {pages} {193202} (\bibinfo {year}
  {2008})}\BibitemShut {NoStop}%
\bibitem [{SMA()}]{SMAMR}%
  \BibitemOpen
  \href@noop {} {\bibinfo  {journal} {See Supplemental Material for details of
  the effective models and of calculations of Berry curvature, orbital magnetic
  moments and anisotropic conductivity of bulk states and surface states. The
  Supplemental Material also contains
  Refs.~\cite{shen2013TIDirac,zhang2009NP,Chang2008jpcm,EPhonon2001Ziman,Sundaram1999PRB,Xiao2010RMP,Shan2010njp,taskin_planar_2017,Chiba2017PRB,Fuseya2015jpsj,YangSY2020PRR,ZhangDQ2019PRL,Wu2022NL}}\
  }\BibitemShut {NoStop}%
\bibitem [{InE()}]{InEx}%
  \BibitemOpen
\bibfield  {journal} {  }\href@noop {} {\bibinfo  {journal} {The terminology
  intrinsic here implies that the anisotropic magnetoresistance is determined
  by the fundamental intrinsic quantum geometric quantities of electrons: Berry
  curvature and orbital magnetic moments}\ }\BibitemShut {NoStop}%
\bibitem [{\citenamefont {Shtrikman}\ and\ \citenamefont
  {Thomas}(1965)}]{SHTRIKMAN1965SSC}%
  \BibitemOpen
\bibfield  {journal} {  }\bibfield  {author} {\bibinfo {author} {\bibfnamefont
  {S.}~\bibnamefont {Shtrikman}}\ and\ \bibinfo {author} {\bibfnamefont
  {H.}~\bibnamefont {Thomas}},\ }\bibfield  {title} {\bibinfo {title} {{Remarks
  on linear magneto-resistance and magneto-heat-conductivity}},\ }\href
  {https://doi.org/https://doi.org/10.1016/0038-1098(65)90178-X} {\bibfield
  {journal} {\bibinfo  {journal} {Solid State Commun.}\ }\textbf {\bibinfo
  {volume} {3}},\ \bibinfo {pages} {147} (\bibinfo {year} {1965})}\BibitemShut
  {NoStop}%
\bibitem [{\citenamefont {Cullen}\ \emph {et~al.}(2021)\citenamefont {Cullen},
  \citenamefont {Bhalla}, \citenamefont {Marcellina}, \citenamefont
  {Hamilton},\ and\ \citenamefont {Culcer}}]{Cullen2021PRL}%
  \BibitemOpen
  \bibfield  {author} {\bibinfo {author} {\bibfnamefont {J.~H.}\ \bibnamefont
  {Cullen}}, \bibinfo {author} {\bibfnamefont {P.}~\bibnamefont {Bhalla}},
  \bibinfo {author} {\bibfnamefont {E.}~\bibnamefont {Marcellina}}, \bibinfo
  {author} {\bibfnamefont {A.~R.}\ \bibnamefont {Hamilton}},\ and\ \bibinfo
  {author} {\bibfnamefont {D.}~\bibnamefont {Culcer}},\ }\bibfield  {title}
  {\bibinfo {title} {{Generating a Topological Anomalous Hall Effect in a
  Nonmagnetic Conductor: An In-Plane Magnetic Field as a Direct Probe of the
  Berry Curvature}},\ }\href {https://doi.org/10.1103/PhysRevLett.126.256601}
  {\bibfield  {journal} {\bibinfo  {journal} {Phys. Rev. Lett.}\ }\textbf
  {\bibinfo {volume} {126}},\ \bibinfo {pages} {256601} (\bibinfo {year}
  {2021})}\BibitemShut {NoStop}%
\bibitem [{\citenamefont {Cao}\ \emph {et~al.}(2023)\citenamefont {Cao},
  \citenamefont {Jiang}, \citenamefont {Li}, \citenamefont {Tu}, \citenamefont
  {Zhou}, \citenamefont {Zhou},\ and\ \citenamefont {Yao}}]{CaoJ2023PRL}%
  \BibitemOpen
  \bibfield  {author} {\bibinfo {author} {\bibfnamefont {J.}~\bibnamefont
  {Cao}}, \bibinfo {author} {\bibfnamefont {W.}~\bibnamefont {Jiang}}, \bibinfo
  {author} {\bibfnamefont {X.-P.}\ \bibnamefont {Li}}, \bibinfo {author}
  {\bibfnamefont {D.}~\bibnamefont {Tu}}, \bibinfo {author} {\bibfnamefont
  {J.}~\bibnamefont {Zhou}}, \bibinfo {author} {\bibfnamefont {J.}~\bibnamefont
  {Zhou}},\ and\ \bibinfo {author} {\bibfnamefont {Y.}~\bibnamefont {Yao}},\
  }\bibfield  {title} {\bibinfo {title} {{In-Plane Anomalous Hall Effect in
  $\mathcal{PT}$-Symmetric Antiferromagnetic Materials}},\ }\href
  {https://doi.org/10.1103/PhysRevLett.130.166702} {\bibfield  {journal}
  {\bibinfo  {journal} {Phys. Rev. Lett.}\ }\textbf {\bibinfo {volume} {130}},\
  \bibinfo {pages} {166702} (\bibinfo {year} {2023})}\BibitemShut {NoStop}%
\bibitem [{\citenamefont {Wang}\ \emph {et~al.}(2024)\citenamefont {Wang},
  \citenamefont {Huang}, \citenamefont {Liu}, \citenamefont {Feng},
  \citenamefont {Zhu}, \citenamefont {Wu}, \citenamefont {Xiao},\ and\
  \citenamefont {Yang}}]{WangH2024PRL}%
  \BibitemOpen
  \bibfield  {author} {\bibinfo {author} {\bibfnamefont {H.}~\bibnamefont
  {Wang}}, \bibinfo {author} {\bibfnamefont {Y.-X.}\ \bibnamefont {Huang}},
  \bibinfo {author} {\bibfnamefont {H.}~\bibnamefont {Liu}}, \bibinfo {author}
  {\bibfnamefont {X.}~\bibnamefont {Feng}}, \bibinfo {author} {\bibfnamefont
  {J.}~\bibnamefont {Zhu}}, \bibinfo {author} {\bibfnamefont {W.}~\bibnamefont
  {Wu}}, \bibinfo {author} {\bibfnamefont {C.}~\bibnamefont {Xiao}},\ and\
  \bibinfo {author} {\bibfnamefont {S.~A.}\ \bibnamefont {Yang}},\ }\bibfield
  {title} {\bibinfo {title} {{Orbital Origin of the Intrinsic Planar Hall
  Effect}},\ }\href {https://doi.org/10.1103/PhysRevLett.132.056301} {\bibfield
   {journal} {\bibinfo  {journal} {Phys. Rev. Lett.}\ }\textbf {\bibinfo
  {volume} {132}},\ \bibinfo {pages} {056301} (\bibinfo {year}
  {2024})}\BibitemShut {NoStop}%
\bibitem [{\citenamefont {Gygax}\ \emph {et~al.}(1986)\citenamefont {Gygax},
  \citenamefont {Schenck}, \citenamefont {van~der Wal},\ and\ \citenamefont
  {Barth}}]{Gygax1986PRL}%
  \BibitemOpen
  \bibfield  {author} {\bibinfo {author} {\bibfnamefont {F.~N.}\ \bibnamefont
  {Gygax}}, \bibinfo {author} {\bibfnamefont {A.}~\bibnamefont {Schenck}},
  \bibinfo {author} {\bibfnamefont {A.~J.}\ \bibnamefont {van~der Wal}},\ and\
  \bibinfo {author} {\bibfnamefont {S.}~\bibnamefont {Barth}},\ }\bibfield
  {title} {\bibinfo {title} {{Higher-Order Angular Dependence of the
  Positive-Muon Knight Shift in Bismuth}},\ }\href
  {https://doi.org/10.1103/PhysRevLett.56.2842} {\bibfield  {journal} {\bibinfo
   {journal} {Phys. Rev. Lett.}\ }\textbf {\bibinfo {volume} {56}},\ \bibinfo
  {pages} {2842} (\bibinfo {year} {1986})}\BibitemShut {NoStop}%
\bibitem [{\citenamefont {Collaudin}\ \emph {et~al.}(2015)\citenamefont
  {Collaudin}, \citenamefont {Fauqu\'e}, \citenamefont {Fuseya}, \citenamefont
  {Kang},\ and\ \citenamefont {Behnia}}]{Collaudin2015PRX}%
  \BibitemOpen
  \bibfield  {author} {\bibinfo {author} {\bibfnamefont {A.}~\bibnamefont
  {Collaudin}}, \bibinfo {author} {\bibfnamefont {B.}~\bibnamefont {Fauqu\'e}},
  \bibinfo {author} {\bibfnamefont {Y.}~\bibnamefont {Fuseya}}, \bibinfo
  {author} {\bibfnamefont {W.}~\bibnamefont {Kang}},\ and\ \bibinfo {author}
  {\bibfnamefont {K.}~\bibnamefont {Behnia}},\ }\bibfield  {title} {\bibinfo
  {title} {{Angle Dependence of the Orbital Magnetoresistance in Bismuth}},\
  }\href {https://doi.org/10.1103/PhysRevX.5.021022} {\bibfield  {journal}
  {\bibinfo  {journal} {Phys. Rev. X}\ }\textbf {\bibinfo {volume} {5}},\
  \bibinfo {pages} {021022} (\bibinfo {year} {2015})}\BibitemShut {NoStop}%
\bibitem [{\citenamefont {Feldman}\ \emph {et~al.}(2016)\citenamefont
  {Feldman}, \citenamefont {Randeria}, \citenamefont {Gyenis}, \citenamefont
  {Wu}, \citenamefont {Ji}, \citenamefont {Cava}, \citenamefont {MacDonald},\
  and\ \citenamefont {Yazdani}}]{Feldman2016Science}%
  \BibitemOpen
  \bibfield  {author} {\bibinfo {author} {\bibfnamefont {B.~E.}\ \bibnamefont
  {Feldman}}, \bibinfo {author} {\bibfnamefont {M.~T.}\ \bibnamefont
  {Randeria}}, \bibinfo {author} {\bibfnamefont {A.}~\bibnamefont {Gyenis}},
  \bibinfo {author} {\bibfnamefont {F.}~\bibnamefont {Wu}}, \bibinfo {author}
  {\bibfnamefont {H.}~\bibnamefont {Ji}}, \bibinfo {author} {\bibfnamefont
  {R.~J.}\ \bibnamefont {Cava}}, \bibinfo {author} {\bibfnamefont {A.~H.}\
  \bibnamefont {MacDonald}},\ and\ \bibinfo {author} {\bibfnamefont
  {A.}~\bibnamefont {Yazdani}},\ }\bibfield  {title} {\bibinfo {title}
  {{Observation of a nematic quantum Hall liquid on the surface of bismuth}},\
  }\href {https://doi.org/10.1126/science.aag1715} {\bibfield  {journal}
  {\bibinfo  {journal} {Science}\ }\textbf {\bibinfo {volume} {354}},\ \bibinfo
  {pages} {316} (\bibinfo {year} {2016})}\BibitemShut {NoStop}%
\bibitem [{\citenamefont {Ito}\ \emph {et~al.}(2016)\citenamefont {Ito},
  \citenamefont {Feng}, \citenamefont {Arita}, \citenamefont {Takayama},
  \citenamefont {Liu}, \citenamefont {Someya}, \citenamefont {Chen},
  \citenamefont {Iimori}, \citenamefont {Namatame}, \citenamefont {Taniguchi},
  \citenamefont {Cheng}, \citenamefont {Tang}, \citenamefont {Komori},
  \citenamefont {Kobayashi}, \citenamefont {Chiang},\ and\ \citenamefont
  {Matsuda}}]{Ito2016PRL}%
  \BibitemOpen
  \bibfield  {author} {\bibinfo {author} {\bibfnamefont {S.}~\bibnamefont
  {Ito}}, \bibinfo {author} {\bibfnamefont {B.}~\bibnamefont {Feng}}, \bibinfo
  {author} {\bibfnamefont {M.}~\bibnamefont {Arita}}, \bibinfo {author}
  {\bibfnamefont {A.}~\bibnamefont {Takayama}}, \bibinfo {author}
  {\bibfnamefont {R.-Y.}\ \bibnamefont {Liu}}, \bibinfo {author} {\bibfnamefont
  {T.}~\bibnamefont {Someya}}, \bibinfo {author} {\bibfnamefont {W.-C.}\
  \bibnamefont {Chen}}, \bibinfo {author} {\bibfnamefont {T.}~\bibnamefont
  {Iimori}}, \bibinfo {author} {\bibfnamefont {H.}~\bibnamefont {Namatame}},
  \bibinfo {author} {\bibfnamefont {M.}~\bibnamefont {Taniguchi}}, \bibinfo
  {author} {\bibfnamefont {C.-M.}\ \bibnamefont {Cheng}}, \bibinfo {author}
  {\bibfnamefont {S.-J.}\ \bibnamefont {Tang}}, \bibinfo {author}
  {\bibfnamefont {F.}~\bibnamefont {Komori}}, \bibinfo {author} {\bibfnamefont
  {K.}~\bibnamefont {Kobayashi}}, \bibinfo {author} {\bibfnamefont {T.-C.}\
  \bibnamefont {Chiang}},\ and\ \bibinfo {author} {\bibfnamefont
  {I.}~\bibnamefont {Matsuda}},\ }\bibfield  {title} {\bibinfo {title}
  {{Proving Nontrivial Topology of Pure Bismuth by Quantum Confinement}},\
  }\href {https://doi.org/10.1103/PhysRevLett.117.236402} {\bibfield  {journal}
  {\bibinfo  {journal} {Phys. Rev. Lett.}\ }\textbf {\bibinfo {volume} {117}},\
  \bibinfo {pages} {236402} (\bibinfo {year} {2016})}\BibitemShut {NoStop}%
\bibitem [{\citenamefont {Zhu}\ \emph {et~al.}(2018)\citenamefont {Zhu},
  \citenamefont {Fauqu\'e}, \citenamefont {Behnia},\ and\ \citenamefont
  {Fuseya}}]{Zhu2018jpcm}%
  \BibitemOpen
  \bibfield  {author} {\bibinfo {author} {\bibfnamefont {Z.}~\bibnamefont
  {Zhu}}, \bibinfo {author} {\bibfnamefont {B.}~\bibnamefont {Fauqu\'e}},
  \bibinfo {author} {\bibfnamefont {K.}~\bibnamefont {Behnia}},\ and\ \bibinfo
  {author} {\bibfnamefont {Y.}~\bibnamefont {Fuseya}},\ }\bibfield  {title}
  {\bibinfo {title} {{Magnetoresistance and valley degree of freedom in bulk
  bismuth}},\ }\href {https://doi.org/10.1088/1361-648X/aaced7} {\bibfield
  {journal} {\bibinfo  {journal} {Journal of Physics: Condensed Matter}\
  }\textbf {\bibinfo {volume} {30}},\ \bibinfo {pages} {313001} (\bibinfo
  {year} {2018})}\BibitemShut {NoStop}%
\bibitem [{\citenamefont {Schindler}\ \emph {et~al.}(2018)\citenamefont
  {Schindler}, \citenamefont {Wang}, \citenamefont {Vergniory}, \citenamefont
  {Cook}, \citenamefont {Murani}, \citenamefont {Sengupta}, \citenamefont
  {Kasumov}, \citenamefont {Deblock}, \citenamefont {Jeon}, \citenamefont
  {Drozdov}, \citenamefont {Bouchiat}, \citenamefont {Gu{\'e}ron},
  \citenamefont {Yazdani}, \citenamefont {Bernevig},\ and\ \citenamefont
  {Neupert}}]{Schindler2018NP}%
  \BibitemOpen
  \bibfield  {author} {\bibinfo {author} {\bibfnamefont {F.}~\bibnamefont
  {Schindler}}, \bibinfo {author} {\bibfnamefont {Z.}~\bibnamefont {Wang}},
  \bibinfo {author} {\bibfnamefont {M.~G.}\ \bibnamefont {Vergniory}}, \bibinfo
  {author} {\bibfnamefont {A.~M.}\ \bibnamefont {Cook}}, \bibinfo {author}
  {\bibfnamefont {A.}~\bibnamefont {Murani}}, \bibinfo {author} {\bibfnamefont
  {S.}~\bibnamefont {Sengupta}}, \bibinfo {author} {\bibfnamefont {A.~Y.}\
  \bibnamefont {Kasumov}}, \bibinfo {author} {\bibfnamefont {R.}~\bibnamefont
  {Deblock}}, \bibinfo {author} {\bibfnamefont {S.}~\bibnamefont {Jeon}},
  \bibinfo {author} {\bibfnamefont {I.}~\bibnamefont {Drozdov}}, \bibinfo
  {author} {\bibfnamefont {H.}~\bibnamefont {Bouchiat}}, \bibinfo {author}
  {\bibfnamefont {S.}~\bibnamefont {Gu{\'e}ron}}, \bibinfo {author}
  {\bibfnamefont {A.}~\bibnamefont {Yazdani}}, \bibinfo {author} {\bibfnamefont
  {B.~A.}\ \bibnamefont {Bernevig}},\ and\ \bibinfo {author} {\bibfnamefont
  {T.}~\bibnamefont {Neupert}},\ }\bibfield  {title} {\bibinfo {title}
  {{Higher-order topology in bismuth}},\ }\href
  {https://doi.org/10.1038/s41567-018-0224-7} {\bibfield  {journal} {\bibinfo
  {journal} {Nat. Phys.}\ }\textbf {\bibinfo {volume} {14}},\ \bibinfo {pages}
  {918} (\bibinfo {year} {2018})}\BibitemShut {NoStop}%
\bibitem [{\citenamefont {Aggarwal}\ \emph {et~al.}(2021)\citenamefont
  {Aggarwal}, \citenamefont {Zhu}, \citenamefont {Hughes},\ and\ \citenamefont
  {Madhavan}}]{Aggarwal2021NC}%
  \BibitemOpen
  \bibfield  {author} {\bibinfo {author} {\bibfnamefont {L.}~\bibnamefont
  {Aggarwal}}, \bibinfo {author} {\bibfnamefont {P.}~\bibnamefont {Zhu}},
  \bibinfo {author} {\bibfnamefont {T.~L.}\ \bibnamefont {Hughes}},\ and\
  \bibinfo {author} {\bibfnamefont {V.}~\bibnamefont {Madhavan}},\ }\bibfield
  {title} {\bibinfo {title} {{Evidence for higher order topology in Bi and
  Bi${}_{0.92}$Sbi${}_{0.08}$}},\ }\href
  {https://doi.org/10.1038/s41467-021-24683-8} {\bibfield  {journal} {\bibinfo
  {journal} {Nature Communications}\ }\textbf {\bibinfo {volume} {12}},\
  \bibinfo {pages} {4420} (\bibinfo {year} {2021})}\BibitemShut {NoStop}%
\bibitem [{\citenamefont {Liu}\ and\ \citenamefont
  {Allen}(1995)}]{LiuY1995PRB}%
  \BibitemOpen
  \bibfield  {author} {\bibinfo {author} {\bibfnamefont {Y.}~\bibnamefont
  {Liu}}\ and\ \bibinfo {author} {\bibfnamefont {R.~E.}\ \bibnamefont
  {Allen}},\ }\bibfield  {title} {\bibinfo {title} {{Electronic structure of
  the semimetals Bi and Sb}},\ }\href
  {https://doi.org/10.1103/PhysRevB.52.1566} {\bibfield  {journal} {\bibinfo
  {journal} {Phys. Rev. B}\ }\textbf {\bibinfo {volume} {52}},\ \bibinfo
  {pages} {1566} (\bibinfo {year} {1995})}\BibitemShut {NoStop}%
\bibitem [{\citenamefont {Wolff}(1964)}]{WOLFF1964jpcs}%
  \BibitemOpen
  \bibfield  {author} {\bibinfo {author} {\bibfnamefont {P.}~\bibnamefont
  {Wolff}},\ }\bibfield  {title} {\bibinfo {title} {{Matrix elements and
  selection rules for the two-band model of bismuth}},\ }\href
  {https://doi.org/https://doi.org/10.1016/0022-3697(64)90128-3} {\bibfield
  {journal} {\bibinfo  {journal} {Journal of Physics and Chemistry of Solids}\
  }\textbf {\bibinfo {volume} {25}},\ \bibinfo {pages} {1057} (\bibinfo {year}
  {1964})}\BibitemShut {NoStop}%
\bibitem [{\citenamefont {Fuseya}\ \emph
  {et~al.}(2015{\natexlab{a}})\citenamefont {Fuseya}, \citenamefont {Zhu},
  \citenamefont {Fauqu\'e}, \citenamefont {Kang}, \citenamefont {Lenoir},\ and\
  \citenamefont {Behnia}}]{Fuseya2015PRL}%
  \BibitemOpen
  \bibfield  {author} {\bibinfo {author} {\bibfnamefont {Y.}~\bibnamefont
  {Fuseya}}, \bibinfo {author} {\bibfnamefont {Z.}~\bibnamefont {Zhu}},
  \bibinfo {author} {\bibfnamefont {B.}~\bibnamefont {Fauqu\'e}}, \bibinfo
  {author} {\bibfnamefont {W.}~\bibnamefont {Kang}}, \bibinfo {author}
  {\bibfnamefont {B.}~\bibnamefont {Lenoir}},\ and\ \bibinfo {author}
  {\bibfnamefont {K.}~\bibnamefont {Behnia}},\ }\bibfield  {title} {\bibinfo
  {title} {{Origin of the Large Anisotropic $g$ Factor of Holes in Bismuth}},\
  }\href {https://doi.org/10.1103/PhysRevLett.115.216401} {\bibfield  {journal}
  {\bibinfo  {journal} {Phys. Rev. Lett.}\ }\textbf {\bibinfo {volume} {115}},\
  \bibinfo {pages} {216401} (\bibinfo {year} {2015}{\natexlab{a}})}\BibitemShut
  {NoStop}%
\bibitem [{Bi4()}]{Bi4amr}%
  \BibitemOpen
  \href@noop {} {\bibinfo  {journal} {Note that due to the absence of the
  electron or hole pockets with fourfold symmetry in the plane, the previous
  mechanism that relay on special crystal symmetry becomes
  irrelevant~\cite{Yahagi2020jpsj}}\ }\BibitemShut {NoStop}%
\bibitem [{dat()}]{dataBi}%
  \BibitemOpen
\bibfield  {journal} {  }\href@noop {} {\bibinfo  {journal} {We extract the
  experimental data of the AMR of Bismuth from the Figs. S5-6 of
  \citep{YangSY2020PRR}. We then fit these data with an empirical formula
  $a+b\mathrm{cos\;}2\theta+c\mathrm{cos\;}4\theta$, assign $b/a$ and $c/a$ as
  the relative magnitude of twofold and fourfold AMRs}\ }\BibitemShut {NoStop}%
\bibitem [{\citenamefont {Bansil}\ \emph {et~al.}(2016)\citenamefont {Bansil},
  \citenamefont {Lin},\ and\ \citenamefont {Das}}]{Bansil2016RMP}%
  \BibitemOpen
\bibfield  {journal} {  }\bibfield  {author} {\bibinfo {author} {\bibfnamefont
  {A.}~\bibnamefont {Bansil}}, \bibinfo {author} {\bibfnamefont
  {H.}~\bibnamefont {Lin}},\ and\ \bibinfo {author} {\bibfnamefont
  {T.}~\bibnamefont {Das}},\ }\bibfield  {title} {\bibinfo {title}
  {{Topological band theory}},\ }\href
  {https://doi.org/10.1103/RevModPhys.88.021004} {\bibfield  {journal}
  {\bibinfo  {journal} {Rev. Mod. Phys.}\ }\textbf {\bibinfo {volume} {88}},\
  \bibinfo {pages} {021004} (\bibinfo {year} {2016})}\BibitemShut {NoStop}%
\bibitem [{\citenamefont {Zhang}\ \emph {et~al.}(2009)\citenamefont {Zhang},
  \citenamefont {Liu}, \citenamefont {Qi}, \citenamefont {Dai}, \citenamefont
  {Fang},\ and\ \citenamefont {Zhang}}]{zhang2009NP}%
  \BibitemOpen
  \bibfield  {author} {\bibinfo {author} {\bibfnamefont {H.}~\bibnamefont
  {Zhang}}, \bibinfo {author} {\bibfnamefont {C.-X.}\ \bibnamefont {Liu}},
  \bibinfo {author} {\bibfnamefont {X.-L.}\ \bibnamefont {Qi}}, \bibinfo
  {author} {\bibfnamefont {X.}~\bibnamefont {Dai}}, \bibinfo {author}
  {\bibfnamefont {Z.}~\bibnamefont {Fang}},\ and\ \bibinfo {author}
  {\bibfnamefont {S.-C.}\ \bibnamefont {Zhang}},\ }\bibfield  {title} {\bibinfo
  {title} {{Topological insulators in Bi${}_2$Se${}_3$, Bi${}_2$Te${}_3$ and
  Sb${}_2$Te${}_3$ with a single Dirac cone on the surface}},\ }\href
  {https://doi.org/10.1038/nphys1270} {\bibfield  {journal} {\bibinfo
  {journal} {Nature. Phys}\ }\textbf {\bibinfo {volume} {5}},\ \bibinfo {pages}
  {438} (\bibinfo {year} {2009})}\BibitemShut {NoStop}%
\bibitem [{\citenamefont {Shan}\ \emph {et~al.}(2010)\citenamefont {Shan},
  \citenamefont {Lu},\ and\ \citenamefont {Shen}}]{Shan2010njp}%
  \BibitemOpen
  \bibfield  {author} {\bibinfo {author} {\bibfnamefont {W.-Y.}\ \bibnamefont
  {Shan}}, \bibinfo {author} {\bibfnamefont {H.-Z.}\ \bibnamefont {Lu}},\ and\
  \bibinfo {author} {\bibfnamefont {S.-Q.}\ \bibnamefont {Shen}},\ }\bibfield
  {title} {\bibinfo {title} {{Effective continuous model for surface states and
  thin films of three-dimensional topological insulators}},\ }\href
  {https://doi.org/10.1088/1367-2630/12/4/043048} {\bibfield  {journal}
  {\bibinfo  {journal} {New Journal of Physics}\ }\textbf {\bibinfo {volume}
  {12}},\ \bibinfo {pages} {043048} (\bibinfo {year} {2010})}\BibitemShut
  {NoStop}%
\bibitem [{\citenamefont {Jiang}\ \emph {et~al.}(2022)\citenamefont {Jiang},
  \citenamefont {Liu},\ and\ \citenamefont {Wang}}]{JiangYD2022PRB}%
  \BibitemOpen
  \bibfield  {author} {\bibinfo {author} {\bibfnamefont {Y.}~\bibnamefont
  {Jiang}}, \bibinfo {author} {\bibfnamefont {Z.}~\bibnamefont {Liu}},\ and\
  \bibinfo {author} {\bibfnamefont {J.}~\bibnamefont {Wang}},\ }\bibfield
  {title} {\bibinfo {title} {{Unoccupied topological surface state in
  ${\mathrm{MnBi}}_{2}{\mathrm{Te}}_{4}$}},\ }\href
  {https://doi.org/10.1103/PhysRevB.106.045148} {\bibfield  {journal} {\bibinfo
   {journal} {Phys. Rev. B}\ }\textbf {\bibinfo {volume} {106}},\ \bibinfo
  {pages} {045148} (\bibinfo {year} {2022})}\BibitemShut {NoStop}%
\bibitem [{\citenamefont {Manchon}\ \emph {et~al.}(2015)\citenamefont
  {Manchon}, \citenamefont {Koo}, \citenamefont {Nitta}, \citenamefont
  {Frolov},\ and\ \citenamefont {Duine}}]{Manchon2015NM}%
  \BibitemOpen
  \bibfield  {author} {\bibinfo {author} {\bibfnamefont {A.}~\bibnamefont
  {Manchon}}, \bibinfo {author} {\bibfnamefont {H.~C.}\ \bibnamefont {Koo}},
  \bibinfo {author} {\bibfnamefont {J.}~\bibnamefont {Nitta}}, \bibinfo
  {author} {\bibfnamefont {S.~M.}\ \bibnamefont {Frolov}},\ and\ \bibinfo
  {author} {\bibfnamefont {R.~A.}\ \bibnamefont {Duine}},\ }\bibfield  {title}
  {\bibinfo {title} {{New perspectives for Rashba spin-orbit coupling}},\
  }\href {http://dx.doi.org/10.1038/nmat4360} {\bibfield  {journal} {\bibinfo
  {journal} {Nat. Mater.}\ }\textbf {\bibinfo {volume} {14}},\ \bibinfo {pages}
  {871} (\bibinfo {year} {2015})}\BibitemShut {NoStop}%
\bibitem [{\citenamefont {Zhang}\ \emph {et~al.}(2021)\citenamefont {Zhang},
  \citenamefont {Li}, \citenamefont {Pe\~na Benitez}, \citenamefont
  {Sur\'owka}, \citenamefont {Moessner}, \citenamefont {Molenkamp},\ and\
  \citenamefont {Trauzettel}}]{ZhangSB2021PRL}%
  \BibitemOpen
  \bibfield  {author} {\bibinfo {author} {\bibfnamefont {S.-B.}\ \bibnamefont
  {Zhang}}, \bibinfo {author} {\bibfnamefont {C.-A.}\ \bibnamefont {Li}},
  \bibinfo {author} {\bibfnamefont {F.}~\bibnamefont {Pe\~na Benitez}},
  \bibinfo {author} {\bibfnamefont {P.}~\bibnamefont {Sur\'owka}}, \bibinfo
  {author} {\bibfnamefont {R.}~\bibnamefont {Moessner}}, \bibinfo {author}
  {\bibfnamefont {L.~W.}\ \bibnamefont {Molenkamp}},\ and\ \bibinfo {author}
  {\bibfnamefont {B.}~\bibnamefont {Trauzettel}},\ }\bibfield  {title}
  {\bibinfo {title} {{Super-Resonant Transport of Topological Surface States
  Subjected to In-Plane Magnetic Fields}},\ }\href
  {https://doi.org/10.1103/PhysRevLett.127.076601} {\bibfield  {journal}
  {\bibinfo  {journal} {Phys. Rev. Lett.}\ }\textbf {\bibinfo {volume} {127}},\
  \bibinfo {pages} {076601} (\bibinfo {year} {2021})}\BibitemShut {NoStop}%
\bibitem [{MBT()}]{MBTParameters}%
  \BibitemOpen
  \href@noop {} {\bibinfo  {journal} {In the numerical calculations, we adopt
  the realistic effective model with anisotropy in velocity and the $k^{2}$
  corrections to the mass~\cite{ZhangDQ2019PRL}. The g-factor is equal to 20}\
  }\BibitemShut {NoStop}%
\bibitem [{\citenamefont {Chiba}\ \emph {et~al.}(2017)\citenamefont {Chiba},
  \citenamefont {Takahashi},\ and\ \citenamefont {Bauer}}]{Chiba2017PRB}%
  \BibitemOpen
\bibfield  {journal} {  }\bibfield  {author} {\bibinfo {author} {\bibfnamefont
  {T.}~\bibnamefont {Chiba}}, \bibinfo {author} {\bibfnamefont
  {S.}~\bibnamefont {Takahashi}},\ and\ \bibinfo {author} {\bibfnamefont
  {G.~E.~W.}\ \bibnamefont {Bauer}},\ }\bibfield  {title} {\bibinfo {title}
  {{Magnetic-proximity-induced magnetoresistance on topological insulators}},\
  }\href {https://doi.org/10.1103/PhysRevB.95.094428} {\bibfield  {journal}
  {\bibinfo  {journal} {Phys. Rev. B}\ }\textbf {\bibinfo {volume} {95}},\
  \bibinfo {pages} {094428} (\bibinfo {year} {2017})}\BibitemShut {NoStop}%
\bibitem [{\citenamefont {V\'yborn\'y}\ \emph {et~al.}(2009)\citenamefont
  {V\'yborn\'y}, \citenamefont {Ku\ifmmode~\check{c}\else \v{c}\fi{}era},
  \citenamefont {Sinova}, \citenamefont {Rushforth}, \citenamefont
  {Gallagher},\ and\ \citenamefont {Jungwirth}}]{Vyborny2009PRB}%
  \BibitemOpen
  \bibfield  {author} {\bibinfo {author} {\bibfnamefont {K.}~\bibnamefont
  {V\'yborn\'y}}, \bibinfo {author} {\bibfnamefont {J.}~\bibnamefont
  {Ku\ifmmode~\check{c}\else \v{c}\fi{}era}}, \bibinfo {author} {\bibfnamefont
  {J.}~\bibnamefont {Sinova}}, \bibinfo {author} {\bibfnamefont {A.~W.}\
  \bibnamefont {Rushforth}}, \bibinfo {author} {\bibfnamefont {B.~L.}\
  \bibnamefont {Gallagher}},\ and\ \bibinfo {author} {\bibfnamefont
  {T.}~\bibnamefont {Jungwirth}},\ }\bibfield  {title} {\bibinfo {title}
  {{Microscopic mechanism of the noncrystalline anisotropic magnetoresistance
  in (Ga,Mn)As}},\ }\href {https://doi.org/10.1103/PhysRevB.80.165204}
  {\bibfield  {journal} {\bibinfo  {journal} {Phys. Rev. B}\ }\textbf {\bibinfo
  {volume} {80}},\ \bibinfo {pages} {165204} (\bibinfo {year}
  {2009})}\BibitemShut {NoStop}%
\bibitem [{\citenamefont {Xu}\ \emph {et~al.}(2019)\citenamefont {Xu},
  \citenamefont {Song}, \citenamefont {Wang}, \citenamefont {Weng},\ and\
  \citenamefont {Dai}}]{XuYF2019PRL}%
  \BibitemOpen
  \bibfield  {author} {\bibinfo {author} {\bibfnamefont {Y.}~\bibnamefont
  {Xu}}, \bibinfo {author} {\bibfnamefont {Z.}~\bibnamefont {Song}}, \bibinfo
  {author} {\bibfnamefont {Z.}~\bibnamefont {Wang}}, \bibinfo {author}
  {\bibfnamefont {H.}~\bibnamefont {Weng}},\ and\ \bibinfo {author}
  {\bibfnamefont {X.}~\bibnamefont {Dai}},\ }\bibfield  {title} {\bibinfo
  {title} {{Higher-Order Topology of the Axion Insulator
  ${\mathrm{EuIn}}_{2}{\mathrm{As}}_{2}$}},\ }\href
  {https://doi.org/10.1103/PhysRevLett.122.256402} {\bibfield  {journal}
  {\bibinfo  {journal} {Phys. Rev. Lett.}\ }\textbf {\bibinfo {volume} {122}},\
  \bibinfo {pages} {256402} (\bibinfo {year} {2019})}\BibitemShut {NoStop}%
\bibitem [{\citenamefont {Zhang}\ \emph
  {et~al.}(2019{\natexlab{a}})\citenamefont {Zhang}, \citenamefont {Shi},
  \citenamefont {Zhu}, \citenamefont {Xing}, \citenamefont {Zhang},\ and\
  \citenamefont {Wang}}]{ZhangDQ2019PRL}%
  \BibitemOpen
  \bibfield  {author} {\bibinfo {author} {\bibfnamefont {D.}~\bibnamefont
  {Zhang}}, \bibinfo {author} {\bibfnamefont {M.}~\bibnamefont {Shi}}, \bibinfo
  {author} {\bibfnamefont {T.}~\bibnamefont {Zhu}}, \bibinfo {author}
  {\bibfnamefont {D.}~\bibnamefont {Xing}}, \bibinfo {author} {\bibfnamefont
  {H.}~\bibnamefont {Zhang}},\ and\ \bibinfo {author} {\bibfnamefont
  {J.}~\bibnamefont {Wang}},\ }\bibfield  {title} {\bibinfo {title}
  {{Topological Axion States in the Magnetic Insulator
  ${\mathrm{MnBi}}_{2}{\mathrm{Te}}_{4}$ with the Quantized Magnetoelectric
  Effect}},\ }\href {https://doi.org/10.1103/PhysRevLett.122.206401} {\bibfield
   {journal} {\bibinfo  {journal} {Phys. Rev. Lett.}\ }\textbf {\bibinfo
  {volume} {122}},\ \bibinfo {pages} {206401} (\bibinfo {year}
  {2019}{\natexlab{a}})}\BibitemShut {NoStop}%
\bibitem [{\citenamefont {Zhang}\ \emph {et~al.}(2020)\citenamefont {Zhang},
  \citenamefont {Wu},\ and\ \citenamefont {Das~Sarma}}]{ZhangRX2020PRL}%
  \BibitemOpen
  \bibfield  {author} {\bibinfo {author} {\bibfnamefont {R.-X.}\ \bibnamefont
  {Zhang}}, \bibinfo {author} {\bibfnamefont {F.}~\bibnamefont {Wu}},\ and\
  \bibinfo {author} {\bibfnamefont {S.}~\bibnamefont {Das~Sarma}},\ }\bibfield
  {title} {\bibinfo {title} {{M\"obius Insulator and Higher-Order Topology in
  ${\mathrm{MnBi}}_{2n}{\mathrm{Te}}_{3n+1}$}},\ }\href
  {https://doi.org/10.1103/PhysRevLett.124.136407} {\bibfield  {journal}
  {\bibinfo  {journal} {Phys. Rev. Lett.}\ }\textbf {\bibinfo {volume} {124}},\
  \bibinfo {pages} {136407} (\bibinfo {year} {2020})}\BibitemShut {NoStop}%
\bibitem [{\citenamefont {Taskin}\ \emph {et~al.}(2016)\citenamefont {Taskin},
  \citenamefont {Legg}, \citenamefont {Yang}, \citenamefont {Sasaki},
  \citenamefont {Kanai}, \citenamefont {Matsumoto}, \citenamefont {Rosch},\
  and\ \citenamefont {Ando}}]{taskin_planar_2017}%
  \BibitemOpen
  \bibfield  {author} {\bibinfo {author} {\bibfnamefont {A.~A.}\ \bibnamefont
  {Taskin}}, \bibinfo {author} {\bibfnamefont {H.~F.}\ \bibnamefont {Legg}},
  \bibinfo {author} {\bibfnamefont {F.}~\bibnamefont {Yang}}, \bibinfo {author}
  {\bibfnamefont {S.}~\bibnamefont {Sasaki}}, \bibinfo {author} {\bibfnamefont
  {Y.}~\bibnamefont {Kanai}}, \bibinfo {author} {\bibfnamefont
  {K.}~\bibnamefont {Matsumoto}}, \bibinfo {author} {\bibfnamefont
  {A.}~\bibnamefont {Rosch}},\ and\ \bibinfo {author} {\bibfnamefont
  {Y.}~\bibnamefont {Ando}},\ }\bibfield  {title} {\bibinfo {title} {{Planar
  Hall effect from the surface of topological insulators}},\ }\href
  {https://doi.org/10.1038/s41467-017-01474-8} {\bibfield  {journal} {\bibinfo
  {journal} {Nat. Commun}\ }\textbf {\bibinfo {volume} {8}},\ \bibinfo {pages}
  {1340} (\bibinfo {year} {2016})}\BibitemShut {NoStop}%
\bibitem [{\citenamefont {Zheng}\ \emph {et~al.}(2020)\citenamefont {Zheng},
  \citenamefont {Duan}, \citenamefont {Wang}, \citenamefont {Li}, \citenamefont
  {Deng},\ and\ \citenamefont {Wang}}]{ZhengSH2020PRB}%
  \BibitemOpen
  \bibfield  {author} {\bibinfo {author} {\bibfnamefont {S.-H.}\ \bibnamefont
  {Zheng}}, \bibinfo {author} {\bibfnamefont {H.-J.}\ \bibnamefont {Duan}},
  \bibinfo {author} {\bibfnamefont {J.-K.}\ \bibnamefont {Wang}}, \bibinfo
  {author} {\bibfnamefont {J.-Y.}\ \bibnamefont {Li}}, \bibinfo {author}
  {\bibfnamefont {M.-X.}\ \bibnamefont {Deng}},\ and\ \bibinfo {author}
  {\bibfnamefont {R.-Q.}\ \bibnamefont {Wang}},\ }\bibfield  {title} {\bibinfo
  {title} {{Origin of planar Hall effect on the surface of topological
  insulators: Tilt of Dirac cone by an in-plane magnetic field}},\ }\href
  {https://doi.org/10.1103/PhysRevB.101.041408} {\bibfield  {journal} {\bibinfo
   {journal} {Phys. Rev. B}\ }\textbf {\bibinfo {volume} {101}},\ \bibinfo
  {pages} {041408} (\bibinfo {year} {2020})}\BibitemShut {NoStop}%
\bibitem [{\citenamefont {He}\ \emph {et~al.}(2019)\citenamefont {He},
  \citenamefont {Zhang}, \citenamefont {Zhu}, \citenamefont {Shi},
  \citenamefont {Heinonen}, \citenamefont {Vignale},\ and\ \citenamefont
  {Yang}}]{HeP2019PRL}%
  \BibitemOpen
  \bibfield  {author} {\bibinfo {author} {\bibfnamefont {P.}~\bibnamefont
  {He}}, \bibinfo {author} {\bibfnamefont {S.~S.-L.}\ \bibnamefont {Zhang}},
  \bibinfo {author} {\bibfnamefont {D.}~\bibnamefont {Zhu}}, \bibinfo {author}
  {\bibfnamefont {S.}~\bibnamefont {Shi}}, \bibinfo {author} {\bibfnamefont
  {O.~G.}\ \bibnamefont {Heinonen}}, \bibinfo {author} {\bibfnamefont
  {G.}~\bibnamefont {Vignale}},\ and\ \bibinfo {author} {\bibfnamefont
  {H.}~\bibnamefont {Yang}},\ }\bibfield  {title} {\bibinfo {title} {{Nonlinear
  Planar Hall Effect}},\ }\href
  {https://doi.org/10.1103/PhysRevLett.123.016801} {\bibfield  {journal}
  {\bibinfo  {journal} {Phys. Rev. Lett.}\ }\textbf {\bibinfo {volume} {123}},\
  \bibinfo {pages} {016801} (\bibinfo {year} {2019})}\BibitemShut {NoStop}%
\bibitem [{\citenamefont {Zhang}\ \emph
  {et~al.}(2019{\natexlab{b}})\citenamefont {Zhang}, \citenamefont {Wu},
  \citenamefont {Liu},\ and\ \citenamefont {Yazyev}}]{ZhangSN2019PRB}%
  \BibitemOpen
  \bibfield  {author} {\bibinfo {author} {\bibfnamefont {S.}~\bibnamefont
  {Zhang}}, \bibinfo {author} {\bibfnamefont {Q.}~\bibnamefont {Wu}}, \bibinfo
  {author} {\bibfnamefont {Y.}~\bibnamefont {Liu}},\ and\ \bibinfo {author}
  {\bibfnamefont {O.~V.}\ \bibnamefont {Yazyev}},\ }\bibfield  {title}
  {\bibinfo {title} {{Magnetoresistance from Fermi surface topology}},\ }\href
  {https://doi.org/10.1103/PhysRevB.99.035142} {\bibfield  {journal} {\bibinfo
  {journal} {Phys. Rev. B}\ }\textbf {\bibinfo {volume} {99}},\ \bibinfo
  {pages} {035142} (\bibinfo {year} {2019}{\natexlab{b}})}\BibitemShut
  {NoStop}%
\bibitem [{\citenamefont {Fuseya}\ \emph
  {et~al.}(2015{\natexlab{b}})\citenamefont {Fuseya}, \citenamefont {Ogata},\
  and\ \citenamefont {Fukuyama}}]{Fuseya2015jpsj}%
  \BibitemOpen
  \bibfield  {author} {\bibinfo {author} {\bibfnamefont {Y.}~\bibnamefont
  {Fuseya}}, \bibinfo {author} {\bibfnamefont {M.}~\bibnamefont {Ogata}},\ and\
  \bibinfo {author} {\bibfnamefont {H.}~\bibnamefont {Fukuyama}},\ }\bibfield
  {title} {\bibinfo {title} {{Transport Properties and Diamagnetism of Dirac
  Electrons in Bismuth}},\ }\href {https://doi.org/10.7566/JPSJ.84.012001}
  {\bibfield  {journal} {\bibinfo  {journal} {Journal of the Physical Society
  of Japan}\ }\textbf {\bibinfo {volume} {84}},\ \bibinfo {pages} {012001}
  (\bibinfo {year} {2015}{\natexlab{b}})}\BibitemShut {NoStop}%
\end{thebibliography}%

\end{document}